\documentclass[10pt,onecolumn]{IEEEtran}

\usepackage{makeidx}
\usepackage{amsmath}
\usepackage{epsfig}
\usepackage{amssymb}
\usepackage{textcomp}
\usepackage{pstricks,pst-text,psfrag}
\usepackage{algorithm}
\usepackage{algorithmic}
\usepackage{datetime}
\usepackage{setspace}
\usepackage{watermark}
\usepackage{graphicx}
\usepackage{subfig}

\newcommand{\mbf}{\mathbf}

\newcommand{\bzero}{\mbf{0}}

\def\b1{{\mathbf 1}}

\def\bgamma{{\mbox{\boldmath{$\gamma$}}}}
\def\bGamma{{\mbox{\boldmath{$\Gamma$}}}}
\def\bLambda{{\mbox{\boldmath{$\Lambda$}}}}
\def\blambda{{\mbox{\boldmath{$\lambda$}}}}

\def\bPhi{{\mbox{\boldmath{$\Phi$}}}}
 \def\bPsi{{\mbox{\boldmath{$ \Psi$}}}}

\def\btau{{\mbox{\boldmath{$\tau$}}}}

\def\bSigma{{\mbox{\boldmath{$\Sigma$}}}}

\def\brho{{\mbox{\boldmath{$\rho$}}}}
\def\brhou{\underline{\mbox{\boldmath{$\rho$}}}}

\newcommand{\cN}{\ensuremath{\mathcal{N}}}

\newcommand{\bXu}{\ensuremath{\underline{\bX}}}
\newcommand{\bYu}{\ensuremath{\underline{\bY}}}
\newcommand{\bSu}{\ensuremath{\underline{\bS}}}

\newcommand{\argmin}{\mbox{\textrm arg}\min}

\newcommand{\diag}{\mbox{\textrm diag}}
\newcommand{\bdiag}{\mbox{\textrm bdiag}}

\newcommand{\vect}{\mbox{\textrm vec}}

\def\eg{{e.g.,\ }}
\def\ie{{i.e.,\ }}

\newcommand{\betab}{\begin{tabbing}}
\newcommand{\entab}{\end{tabbing}}
\newcommand{\beitem}{\begin{itemize}}
\newcommand{\enitem}{\end{itemize}}
\newcommand{\bea}{\begin{array}}
\newcommand{\ena}{\end{array}}
\newcommand{\beq}{\begin{equation}}
\newcommand{\enq}{\end{equation}}
\newcommand{\beqa}{\begin{eqnarray}}
\newcommand{\enqa}{\end{eqnarray}}
\newcommand{\beqan}{\begin{eqnarray*}}
\newcommand{\enqan}{\end{eqnarray*}}
\newcommand{\beenum}{\begin{enumerate}}
\newcommand{\enenum}{\end{enumerate}}
\newcommand{\DL}{\begin{dashlist}}
\newcommand{\DLE}{\end{dashlist}}

\newcommand{\ba}{{\ensuremath{\mathbf{a}}}}
\newcommand{\bb}{{\ensuremath{\mathbf{b}}}}

\newcommand{\bd}{{\ensuremath{\mathbf{d}}}}

\newcommand{\bh}{{\ensuremath{\mathbf{h}}}}

\newcommand{\br}{{\ensuremath{\mathbf{r}}}}
\newcommand{\bt}{{\ensuremath{\mathbf{t}}}}

\newcommand{\by}{{\ensuremath{\mathbf{y}}}}
\newcommand{\bz}{{\ensuremath{\mathbf{z}}}}

\newcommand{\bA}{{\ensuremath{\mathbf{A}}}}
\newcommand{\bB}{{\ensuremath{\mathbf{B}}}}
\newcommand{\bC}{{\ensuremath{\mathbf{C}}}}
\newcommand{\bD}{{\ensuremath{\mathbf{D}}}}
\newcommand{\bE}{{\ensuremath{\mathbf{E}}}}
\newcommand{\bF}{{\ensuremath{\mathbf{F}}}}

\newcommand{\bH}{{\ensuremath{\mathbf{H}}}}
\newcommand{\bI}{{\ensuremath{\mathbf{I}}}}
\newcommand{\bJ}{{\ensuremath{\mathbf{J}}}}

\newcommand{\bP}{{\ensuremath{\mathbf{P}}}}

\newcommand{\bR}{{\ensuremath{\mathbf{R}}}}
\newcommand{\bS}{{\ensuremath{\mathbf{S}}}}
\newcommand{\bT}{{\ensuremath{\mathbf{T}}}}
\newcommand{\bU}{{\ensuremath{\mathbf{U}}}}
\newcommand{\bV}{{\ensuremath{\mathbf{V}}}}
\newcommand{\bW}{{\ensuremath{\mathbf{W}}}}
\newcommand{\bX}{{\ensuremath{\mathbf{X}}}}
\newcommand{\bY}{{\ensuremath{\mathbf{Y}}}}

\def\bSigma{{\mbox{\boldmath{$\Sigma$}}}} 
\def\bSigmau{\underline{\mbox{\boldmath{$\Sigma$}}}}

\def\btheta{{\mbox{\boldmath{$\theta$}}}}

\def\btau{{\mbox{\boldmath{$\tau$}}}}

\def\bpsi{{\mbox{\boldmath{$\psi$}}}}

\def\bEta{{\mbox{\boldmath{$\eta$}}}}
\def\bzeta{{\mbox{\boldmath{$\zeta$}}}}

\newcommand{\norm}[1]{\lVert#1\rVert}

\def\ur{{\underline{r}}}
\def\bur{{\underline{\br}}}

\newcommand{\Cramer}{Cram\'{e}r}

\newcommand{\barN}{{\ensuremath{\bar{N}}}}

\newcommand{\byu}{{\ensuremath{\underline{\by}}}}
\newcommand{\bAu}{{\ensuremath{\underline{\bA}}}}
\newcommand{\bBu}{{\ensuremath{\underline{\bB}}}}
\newcommand{\bFu}{{\ensuremath{\underline{\bF}}}}

\newcounter{remarkCounter}
\stepcounter{remarkCounter}
\newcommand{\newRemark}[2]{
{\textbf{Remark \arabic{remarkCounter}}}:
(\textbf{#1}): #2
\stepcounter{remarkCounter}}

\newcommand{\barP}    {\bar{P}}

\newcommand{\dr}    {\dot{r}}
\newcommand{\ddr}   {\ddot{r}}

\newcommand{\bdr}   {\dot{\br}}
\newcommand{\bddr}  {\ddot{\br}}
\newcommand{\bdR}   {\dot{\bR}}
\newcommand{\bddR}  {\ddot{\bR}}

\newcommand{\bdur}  {\dot{\bur}}
\newcommand{\bddur} {\ddot{\bur}}

\newcommand{\firstAuthor}       
{Raj~Thilak~Rajan}
\newcommand{\secondAuthor}      
{Geert~Leus}
\newcommand{\thirdAuthor}       
{Alle-Jan~van~der~Veen}

\newcommand{\theTitle}          {Relative Kinematics of an Anchorless Network}

\title  {\theTitle}
\author {
\firstAuthor, \  \secondAuthor\ and \thirdAuthor
\thanks{Submitted: 27th July 2016 , In revision }
\thanks{A part of this work is published in \cite{rajanThesis2016}}
\thanks{R.T. Rajan, G.J.T. Leus and A.-J. van der Veen are with TU Delft, Delft, The Netherlands (email: r.t.rajan@tudelft.nl; g.j.t.leus@tudelft.nl; a.j.vanderveen@tudelft.nl)}}

\begin{document}
\watermark{\footnotesize Revised manuscript}
\maketitle

\begin{abstract} Estimating the location of $N$ coordinates in a $P$ dimensional Euclidean space from pairwise distances (or proximity measurements), is a principal challenge in a wide variety of fields. Conventionally, when localizing a static network of immobile nodes, non-linear dimensional reduction techniques are applied on the measured Euclidean distance matrix (EDM) to obtain the relative coordinates upto a rotation and translation. In this article, we focus on an anchorless network of mobile nodes, where the distance measurements between the mobile nodes are time-varying in nature. Furthermore, in an anchorless network the absolute knowledge of any node positions, motion or reference frame is absent. We derive a novel data model which relates the time-varying EDMs to the time-varying relative positions of an anchorless network. Using this data model, we estimate the relative position, relative velocity and higher order derivatives, which are collectively termed as the relative kinematics of the anchorless network. The derived data model is inherently ill-posed, however can be solved using certain relative immobility constraints. We propose elegant closed form solutions to recursively estimate the relative kinematics of the network. For the sake of completeness, estimators are also proposed to find the absolute kinematics of the nodes, given known reference anchors. \Cramer-Rao bounds are derived for the new data model and simulations are performed to analyze the performance of the proposed solutions.

\end{abstract}

\section{Introduction} Estimating the relative coordinates of $N$ points (or nodes) in a $P$ dimensional Euclidean space using proximity measurements is a fundamental problem spanning a broad range of applications. These applications include, but are not limited to,  psychometric analysis \cite{koehler2005}, perceptual mapping \cite{ho2008}, range-based anchorless localization \cite{dil2006}, combinatorial-chemistry \cite{agrafiotis2001}, polar-based navigation \cite{rehm2005}, sensor array calibration \cite{jenkins2004} and in general exploratory data analysis \cite{borg97}. In anchorless localization scenarios for instance, nodes heavily rely on co-operative estimation of relative coordinates. Such anchorless networks naturally arise when nodes are inaccessible or only intermittently monitored, as is the case in space-based satellite arrays \cite{rajanJ0}, underwater networks \cite{chandrasekhar2006} or indoor wireless sensor networks \cite{yang2012}. In such reference-free scenarios, the proximity information, often measured as pairwise distances between the nodes, form a key input in estimating the relative coordinates of nodes. These relative coordinates are typically estimated using Non-linear dimensaionality reduction algorithms (such as Multidimensional scaling (MDS)), which have been studied rigorously over the past decades \cite{borg97,young2013}. However, considerably less attention has been directed towards anchorless mobile scenarios.

Our primary focus in this article is on an anchorless network of mobile nodes, where we use the term anchorless to indicate no absolute knowledge of the node positions, motion or reference frame. Furthermore, since the nodes are mobile, both the node positions and the pairwise distance measurements between the nodes are time-varying in nature. Our motive is to relate the time-varying pairwise distance measurements to time-derivatives of the node coordinates. For an anchorless network, these include the relative position, relative velocity, relative acceleration and higher-order derivatives which we cumulatively refer to as relative kinematics in this article. It is worth noting that the universal definition of relative kinematics inherently relies on the information in the absolute reference frame. For example, the non-relativistic relative velocity between two objects is rightly defined as the difference between their respective absolute velocity vectors \cite{halliday2010}. In an anchorless framework however, a natural question arises on whether the relative kinematics can be estimated, given only time-varying distance measurements. Ergo, we wish to understand the relationship between the time-varying distance measurements and the relative kinematics of mobile nodes, which is the prime focus of this article.


\subsection{Previous work}  A key challenge in our pursuit is that both the time-varying distance and the time-varying relative positions are non-linear in nature. In particular, the Euclidean distance between a pair of mobile nodes is almost always a non-linear function of time, even if the nodes are in linear independent motion \cite{rajanJ1}. Therefore, it is perhaps not surprising that traditional methods to solving such a problem have been state-space based approaches with the assistance of known anchors \cite{kay1993}. The initial position of the nodes are estimated using MDS like algorithms, which use the Euclidean distance matrix (EDM) at a single time-instant to estimate the relative node positions. Given this initial estimate, the relative positions are tracked over a period of time with Doppler measurements and known anchors \cite{wei10}, or via subspace tracking methods \cite{jamali2012}. Unfortunately, Doppler measurements and anchor information are not always available. Secondly, subspace tracking is applicable only for small perturbations in motion and therefore offer little insight on the kinematics of the motion itself.


In our previous study, we proposed a two-step solution to estimate relative velocities of the nodes from time-varying distance measurements \cite{rajanJ2}. Firstly, the derivatives of the time-varying distances were estimated by solving a Vandermonde-like system of linear equations. The estimated regression coefficients (called range parameters) jointly yield the relative velocities and the relative positions, using MDS-like algorithms. However, the proposed solution is valid only for linear motion, which is not always practical. Furthermore, the previously proposed MDS-based relative velocity estimator heavily relies on the second-order time-derivative of distance, and under Gaussian noise assumptions, it performs worse than the relative position estimator. Thus, designing more optimal estimators for the relative velocity is one of the key motivations for the pursuit of a generalized framework presented in this article. Moreover, understanding the higher order relative kinematics of motion in Euclidean space via time-varying distance measurements is crucial for next-generation localization technologies.


\subsection{Contributions and overview} We present a novel data model in Section \ref{sec:datamodel}, which relates the time-varying distances to the kinematics of the mobile nodes. More concretely, this relationship is established via the derivatives of the time-varying distance (called range parameters), which is estimated in Section \ref{sec:dynamicRanging} using dynamic ranging. In Section \ref{sec:relativeKinematics} we show that the relationship between the range parameters and the relative kinematics takes the form of a Lyapunov-like set of equations, which is inherently ill-posed. In pursuit of unique solutions, we propose elegant least squares algorithms, which can be solved under certain assumptions. For the sake of completion, in Section \ref{sec:absKinematics}, we also propose similar algorithms for estimating  the absolute kinematics of the nodes, given known reference parameters in the cluster. To compare the performance of our estimators, we derive constrained \Cramer-Rao bounds (CRBs), under Gaussian noise assumption on the data. A optimal choice of the weighting matrix ensures the proposed estimator is the best linear unbiased estimator (BLUE) for the given data model. In addition, unconstrained oracle bounds are also derived in Section \ref{sec:crb}, as a benchmark for next generation estimators. In Section \ref{sec:simulations}, we conduct experiments to validate the performance of the proposed estimators.


\subsection{Notation:} The element-wise matrix Hadamard product is denoted by $\odot$ and $(\cdot)^{\odot N}$ denotes element-wise matrix exponent. The Kronecker product is indicated by $\otimes$, the transpose operator by ($\cdot)^T$ and $\hat{(\cdot)}$ denotes an estimated value. $\b1_N \in \mathbb{R}^{N \times 1}$ is a vector of ones, $\bI_N$ is an $N \times N$ identity matrix, $\bzero_{M,N}$ is an $M \times N$ matrix of zeros and $\| \cdot \|$ is the Euclidean norm. For any vector $\ba$, $\diag(\ba)$ represents a diagonal matrix with $\ba$ on the primary diagonal. For a diagonal matrix $\bA$, $\diag(\bA)$ represents a vector of appropriate length, containing the diagonal elements of the matrix $\bA$. The block diagonal matrix $\bA=\bdiag(\bA_1, \bA_2, \hdots, \bA_N)$ consists of matrices $\bA_1, \bA_2, \hdots, \bA_N$ along the diagonal and zeros elsewhere. The first and second derivatives are indicated by $\dot{(\cdot)}$ and $\ddot{(\cdot)}$ respectively, and more generally the $m$th order derivative is represented by $(\cdot)^{(m)}$. Unless otherwise noted, $\underline{(\cdot)}$ is used to indicate parameters of the relative kinematic model. For matrices of compatible dimensions, we will frequently use the following properties
\begin{eqnarray}
\vect(\bA\bB\bC)
&=&	\big(\bC^T \otimes \bA\big)\vect\big(\bB\big), 
\label{eq:vecABC} \\
\vect(\bA)&=& \bJ \vect(\bA^T),
\label{eq:vecA}
\end{eqnarray} where $\bJ$ is an orthogonal permutation matrix. We define an $N$ dimensional centering matrix as $\bP= \bI_{N} -N^{-1} \b1_N\b1_N^T$. A brief list of frequently used notations are tabulated in \tablename\ \ref{tb:notations}.

\color{black} 
\begin{table}[!t] \small
\renewcommand{\arraystretch}{1.3}
\caption{Notations}
\label{tb:notations}
\centering
\begin{tabular}{l l}
\hline
\bfseries Notation & \bfseries Description\\
\hline\hline
$P$ & Number of dimensions \\
$N$ & Number of nodes ($N > P$) \\
$\bD(t) \in \mathbb{R}^{N \times N}$ & Euclidean distance matrix at time $t$ \\
$\bS(t) \in \mathbb{R}^{P \times N}$ & Absolute positions at time $t$ \\
$\bSu(t) \in \mathbb{R}^{P \times N}$ & Relative positions at time $t$ \\
$\bX \in \mathbb{R}^{P \times N}$ & Absolute instantaneous positions at time $t_0$\\
$\bXu \in \mathbb{R}^{P \times N}$ & Relative instantaneous positions at time $t_0$\\
$\bY_m \in \mathbb{R}^{P \times N}$ & $m$th order absolute  kinematics at $t_0$ \\
$\bYu_m \in \mathbb{R}^{P \times N}$ & $m$th order relative kinematics at $t_0$ \\
$\bH_m \in \mathbb{R}^{P \times P}$ & Rotation matrix of the $m$th order kinematics \\
$\bh_m \in \mathbb{R}^{P \times 1}$ & Translational vector of the $m$th order kinematics \\
\hline
\end{tabular}
\end{table}


\section{Time-varying distances and node kinematics}  \label{sec:datamodel} We begin by modeling the relationship between the time-varying distances, the time-varying positions and the node kinematics. In Section \ref{sec:datamodel:absKinematics}, we expand the time-varying position using a Taylor series, the coefficients of which yield the absolute node kinematics. As an extension, we present a novel relative kinematics model in Section \ref{sec:datamodel:relKinematics}. In Sections \ref{sec:datamodel:distance} and \ref{sec:datamodel:datamodel}, the relationship between the time-varying distances and the node kinematics is derived. Using these definitions, we formalize the problem statement in  Section \ref{sec:datamodel:problemStatement}.

\subsection{Absolute kinematics} \label{sec:datamodel:absKinematics} Consider a cluster of $N$ mobile nodes in a $P$ dimensional Euclidean space ($N > P$), whose positions at time $t$ are given by $\bS(t) \in \mathbb{R}^{P \times N}$. For a small time interval $\Delta t= t - t_0$ around $t_0$, we assume the time-varying position is continuously differentiable $M$ times and that the $M$th derivative exists in the interior of this interval. Therefore, the time-dependent position vectors of the respective nodes can be expanded using a Taylor series,\begin{equation}
    \bS(t) = \bS(t)|_{t=t_0} + \dot{\bS}(t)|_{t=t_0}(t-t_0) + \ddot{\bS}(t)|_{t=t_0} (t-t_0)^2 + \hdots
\end{equation} where $(\bS(t), \dot{\bS}(t), \ddot{\bS}(t), \hdots)$ are the derivatives of the time-varying position vectors. Now let $\bX \triangleq \bS(t)|_{t=t_0}$ be a $P \times N$ matrix containing the initial coordinates of the mobile nodes at time $t=t_0$. Furthermore,  let the instantaneous velocities of the nodes \ie the first-order derivatives of the position vectors $\dot{\bS}(t)|_{t=t_0}$ be denoted by $\bY_1\in \mathbb{R}^{P \times N}$, and in general higher-order derivatives as $\bY_m \ \forall\ 1\le m\le M$. Then, the above equation simplifies to  \begin{equation}
\bS(t)=\bX + {\large\sum^{M}_{m=1}} (m!)^{-1} \bY_m (t-t_0)^{m}\label{eq:absPos}.
\end{equation}


\subsection{Relative kinematics} \label{sec:datamodel:relKinematics} The absolute instantaneous positions at $t=t_0$ are an affine transformation of the relative positions, \ie \begin{eqnarray} 
\bX   &=&\bH_0\bXu +   \bh_0\b1^T_N, \label{eq:relPos} \end{eqnarray} where $\bXu \in \mathbb{R}^{P \times N}$ is the relative position matrix upto a rotation and translation, $\bH_0 \in \mathbb{R}^{P \times P}$ is the unknown rotation and $\bh_0 \in \mathbb{R}^{P \times 1}$ is the unknown translation of the network \cite{borg97}. Now, we extend this well-known relative position definition to the higher-order derivatives. For instance, the velocity of the nodes can be written as \begin{eqnarray}
\bY_1   
&=& \bH_1\tilde{\bYu}_1  +  \bh_1\b1^T_N,\label{eq:relVel}
\end{eqnarray} where $\tilde{\bYu}_1$ represents the instantaneous relative velocities of the network at $t=t_0$. The translational vector $\bh_1$ is the group velocity and $\bH_1$ is the unique rotation matrix of the relative velocities \cite{rajanJ2}. More generally, the $m$th order derivative is an affine model defined as \begin{equation}
\bY_m= \bH_m\tilde{\bYu}_m  +  \bh_m\b1^T_N.
\label{eq:relMthOrder}
\end{equation}


We now define the relative time-varying position as $\bSu(t)= \bH^T_0\bS(t)\bP$, and substituting the affine expressions (\ref{eq:relPos}) and (\ref{eq:relMthOrder}) in (\ref{eq:absPos}) we have \begin{equation}
\bSu(t)= \bXu\bP 
+ {\large\sum^{M}_{m=1}} (m!)^{-1}\bH_0^T\bH_m\tilde{\bYu}_m\bP (t-t_0)^m , \label{eq:relPosX1}
\end{equation} where we exploit the property $\bP\b1_N=\bzero_N$ to eliminate the translation vectors, and enforce the orthonormality of the rotation matrix \ie $\bH_0^T\bH_0=\bI_N$. Observe that the translation vector $\bh_0$ does not affect the above equation. Secondly, for a meaningful interpretation of the relative time-varying position, a reference coordinate system must be chosen \eg $\bH_0= \bI_P$. In summary, without the loss of generality, we assume \begin{equation}  
\bH_0= \bI_P \quad\ \text{and}\ \quad\ \bh_0= \bzero_P.
\label{eq:fixedInitialRef}
\end{equation} and subsequently (\ref{eq:relPosX1}) simplifies to \begin{equation} 
\label{eq:relPosX2} 
\underline{\bS}(t)=
\bXu + {\large\sum^{M}_{m=1}} (m!)^{-1}\bYu_m(t-t_0)^m , 
\end{equation} where we use the following properties \begin{subequations}
\begin{align}
\bXu &= \bXu\bP = \bX\bP, \\
\label{eq:relToAbsMthorder}
\bYu_m &= \bH_m\tilde{\bYu}_m = \bY_m\bP, \\
\label{eq:relTimeVaryingPos}
\underline{\bS}(t)&=\bS(t)\bP.
\end{align}
\end{subequations} Note that (\ref{eq:relPosX2}) represents the relative counterpart of the absolute Taylor expansion (\ref{eq:absPos}), where the $\big(\bXu, \bYu_1, \bYu_2, \hdots, \bYu_M\big)$ denote the relative kinematics of the corresponding absolute kinematics $\big(\bX, \bY_1, \bY_2, \hdots, \bY_M\big)$. Our quest in this article is to estimate the relative and absolute kinematic matrices, given time-varying pairwise distance measurements between the nodes. Consequently, the absolute position $\bS(t)$ and relative position $\bSu(t)$ can then be estimated using (\ref{eq:absPos}) and (\ref{eq:relPosX2}) respectively. 
\color{black} 





\subsection{Time-varying distances} \label{sec:datamodel:distance} Similar to the node positions, the pairwise distances are also time-varying which we denote by the time-varying Euclidean distance matrix (EDM) $\bD(t)\triangleq\ [d_{ij}(t)] \in \mathbb{R}^{N \times N}$ where $d_{ij}(t)$ is the pairwise Euclidean distance between the node pair $(i,j)$ at time instant $t$. More explicitly \begin{eqnarray}
\big(\bD(t)\big)^{\odot 2}= \bzeta(t)\b1^T_N + \b1_N\bzeta^T(t) -2\bS^T(t)\bS(t), \label{eq:EDM}
\end{eqnarray} where $\bzeta(t)= \diag \big(\bS^T(t)\bS(t)\big)$. Observe that $\bD(t)$ is a non-linear function of time $t$, even when the nodes are in independent linear motion and hence $\bD(t)$ is a continuously differentiable function in time. Now, based on the time-varying EDM $\bD(t)$, we define the double centered matrix $\bBu(t)$ \begin{subequations} \label{eq:Bt_1}
\begin{align}
\bBu(t)        
&\triangleq
-0.5\bP\Big(\bD(t) \Big)^{\odot 2}\bP, 
\label{eq:Bt_1a} \\
\intertext{and the time derivatives of the double centered matrix
\Big($\dot{\bBu}(t)$, $\ddot{\bBu}(t)$\Big) for upto $M=2$ as,}
\dot{\bBu}(t)  
&\triangleq 
-\bP\Big( \bD(t) \odot \dot{\bD}(t)\Big)\bP, 
\label{eq:Bt_1b}\\
\ddot{\bBu}(t)  
&\triangleq
-\bP \Big( \bD(t) \odot \ddot{\bD}(t) + (\dot{\bD}(t))^{\odot 2} \Big)\bP, 
\label{eq:Bt_1c}
\end{align} \end{subequations} where $\big(\dot{\bD}(t), \ddot{\bD}(t), \hdots \big)$ are the derivatives of the time-varying EDM, which indicate the radial velocity and other higher-order derivatives. Now, let the EDM and the corresponding derivatives at $t=t_0$ be denoted by $\bD(t)|_{t=t_0}\triangleq\bR= [r_{ij}],\ 
\dot{\bD}(t)|_{t=t_0}\triangleq\dot{\bR}= [\dot{r}_{ij}],\
\ddot{\bD}(t)|_{t=t_0}\triangleq\ddot{\bR}= [\ddot{r}_{ij}], \forall \{i,j\} \le N$ , then with an abuse of notation (\ref{eq:Bt_1}) becomes \begin{subequations} \label{eq:Bt_2} 
\begin{eqnarray} 
\bBu^{(0)}\triangleq \bBu(t) |_{t=t_0}       
&=& -0.5\bP\bR^{\odot 2}\bP, \label{eq:Bt_2a}                    \\
\bBu^{(1)}\triangleq \dot{\bBu}(t)|_{t=t_0}  
&=& -\bP\Big[ \bR \odot \bdR \Big]\bP, \label{eq:Bt_2b}            \\
\bBu^{(2)}\triangleq  \ddot{\bBu}(t)|_{t=t_0} 
&=& -\bP \Big[ \bR\odot\bddR + \bdR^{\odot 2} \Big]\bP, \label{eq:Bt_2c} \end{eqnarray} \end{subequations} and higher-order derivatives can be defined along similar lines. In general, given the distance derivatives at $t_0$, \ie the range parameters $(\bR, \bdR , \bddR, \hdots)$, the double centered matrix $\bBu^{(0)}$ and the corresponding higher-order derivatives $(\bBu^{(1)},\bBu^{(2)}, \hdots)$ can be constructed. In a mobile network, the range parameters may not be readily available, however given all the nodes are capable of two way ranging, the range parameters can be estimated using dynamic ranging \cite{rajanJ1}. 

\subsection{Model}\label{sec:datamodel:datamodel} To understand the relationship between the time-varying distances and the relative kinematics of the nodes, we substitute the definition of the EDM from (\ref{eq:EDM}) in (\ref{eq:Bt_1a}) and differentiate recursively to obtain \begin{subequations} \label{eq:Bt_11}
\begin{eqnarray}
\bBu(t)
&=& 
\bSu^T(t)\bSu(t), 
\label{eq:Bt_11a} \\
\dot{\bBu}(t)
&=& 
\dot{\bSu}^T(t)\bSu(t) + \bSu^T(t)\dot{\bSu}(t), 
\label{eq:Bt_11b}\\
\ddot{\bBu}(t)
&=&
\bSu^T(t)\ddot{\bSu}(t) + \ddot{\bSu}^T(t)\bSu(t) + 2\dot{\bSu}^T(t)\dot{\bSu}(t),
\label{eq:Bt_11c}
\end{eqnarray} \end{subequations} where we use the definition (\ref{eq:relTimeVaryingPos}) and introduce $(\dot{\bSu}(t), \ddot{\bSu}(t), \hdots)$ as the derivatives of $\bSu(t)$. Now, rearranging the terms and substituting the definition of $\bSu(t)$ at $t=t_0$ from (\ref{eq:relPosX2}), we have
\begin{subequations} \label{eq:Bt_3} \begin{align}
\label{eq:Bt_3a}
\bB_{0}\ \triangleq&\
\bBu^{(0)}  			
&=&\ \bXu^T\bXu,\\
\bB_{1}\ \triangleq&\
\bBu^{(1)}  &=&\ \bXu^T\bYu_1 + \bYu_1^T\bXu,	\label{eq:Bt_3b}\\
\bB_{2}\ \triangleq&\
\bBu^{(2)} -2\bYu_1^T\bYu_1 &=&\ \bXu^T\bYu_2 + \bYu_2^T\bXu, \label{eq:Bt_3c}\end{align} \end{subequations} where we introduce the matrices $(\bB_0, \bB_1, \bB_2)$. The joint left and right centering using the centering matrix $\bP$ in (\ref{eq:Bt_1}) ensures that the phase center of the relative kinematic matrices $(\bYu_1, \bYu_2)$ are at $\bzero_P$, similar to the relative position $\bXu$.

\subsubsection{Relative kinematics} Now, for $M=0$, combining (\ref{eq:Bt_2a}) and (\ref{eq:Bt_3a}), we have  \begin{equation} \bB_{0}\ = \bXu^T\bXu = -0.5\bP\bR^{\odot 2}\bP, \label{eq:MDS_datamodel}
\end{equation} and more generally for a given $M\ge 1$, (\ref{eq:Bt_3}) can be generalized to \begin{subequations} 
\label{eq:BM} \begin{eqnarray}  
\bB_{M} 
&\triangleq& \bBu^{(M)} -	\sum^{M-1}_{m=1} \begin{pmatrix} M-1 \\ m \end{pmatrix} 
{\bYu^T_{M-m}}\bYu_m	\label{eq:BM_measurement} \\
&=&  \bXu^T\bYu_{M} + \bYu^T_{M}\bXu, 
\label{eq:BM_rel_unknown} 
\end{eqnarray} \end{subequations} where $\bBu^{(M)}$ is the $M$th derivative of the double centered matrix at $t_0$, which is given by (\ref{eq:Bt_2}) and $\bYu_M$ is the $M$th order relative kinematic matrix.

\newRemark{Measurement matrix $\bB_M$}{We make two critical observations on $\bB_M$ in (\ref{eq:BM_measurement}). \begin{itemize}
    \item Firstly, note that $\bB_M$ is dependent on the range parameters $(\bR, \bdR, \bddR, \hdots)$ via the definition of $\bBu^{(M)}$ (\ref{eq:Bt_2}).
    \item Secondly, $\bB_0\triangleq \bB^{(0)}$ and $\bB_1\triangleq \bB^{(1)}$ can be constructed only based on the range parameters (see (\ref{eq:Bt_3})). However for $M\ge 2$, $\bB_M$ in addition to $\bB^{(M)}$, additionally relies on the relative kinematic matrices of order less than $M$. Hence, if the lower order kinematics $\bY_m \forall 2 \le m < M$ are known, then the measurement matrix $\bB_M$ can be reconstructed.
\end{itemize}}


\subsubsection{Absolute kinematics} In addition to the relative kinematics, (\ref{eq:BM_rel_unknown}) can also be reformulated to estimate the absolute kinematics $\bY_M$ of the network. Recall from (\ref{eq:relToAbsMthorder}), that the relative kinematics of the $M$th order is $\bYu_M=\bY_M\bP$ under the assumption (\ref{eq:fixedInitialRef}). Substituting this expression in (\ref{eq:BM_rel_unknown}), we have 
\begin{equation}
\bB_M=  \bXu^T\bY_M\bP + \bP\bY_M^T\bXu,
\label{eq:BM_abs_unknown} 
\end{equation} which is the absolute kinematic model. 

\subsubsection{Model summary} In summary, if the range parameters $(\bR, \bdR, \bddR, \hdots)$ are available, $\bBu^{(M)}$ can be constructed from (\ref{eq:Bt_2}). Given $\bBu^{(0)}$, we aim to solve for the relative position $\bXu$ using the equation (\ref{eq:MDS_datamodel}), which we use to estimate the higher order kinematics. For $M \ge 1$, the measurement matrix $\bB_M$ can be constructed using $\bBu^{(M)}$ and by substituting the lower order relative kinematic matrices $\bY_m\ \forall\ 2 \le m < M$ in (\ref{eq:BM_measurement}). Finally, given the measurement matrix $\bB_M$ and an estimate of $\bXu$, our goal is to estimate the $M$th order relative kinematics $\bYu_M$ and the absolute kinematics $\bY_M$ for $M \ge 1$, using (\ref{eq:BM_rel_unknown}) and (\ref{eq:BM_abs_unknown}) respectively. We now formulate the problem more concretely in the following section.

\subsection{Problem Statement} \label{sec:datamodel:problemStatement}

\emph{\textbf{Problem statement:}} Given the time-varying pairwise distances $\bD(t)$ between the $N$ nodes in a $P$ dimensional Euclidean space, estimate the relative kinematics ($\bXu, \bYu_1, \bYu_2 \hdots$) and absolute kinematics ($\bY_1, \bY_2 \hdots$) of the mobile network. These estimates subsequently yield the relative (and absolute) time-varying positions.

\emph{\textbf{Solution:}} We propose a two-step solution to the above estimation problem. \begin{itemize}
\item[S1)] \label{p1} \emph{Dynamic ranging and relative position}: Given the time-varying distance measurements $\bD(t)$, we employ dynamic ranging to obtain the range parameters ($\bR, \bdR, \bddR, \hdots$) in Section \ref{sec:dynamicRanging}, under the assumption that all the nodes are capable of communicating with each other. Secondly, we also estimate the initial relative position $\bXu$ using (\ref{eq:MDS_datamodel}).

\item[S2)] \label{p2} \emph{Kinematics}: The measurement matrix $\bB_M$ can be constructed using the estimated range parameters, and lower order kinematics (\ref{eq:BM_measurement}). Given the relative position $\bXu$ and $\bB_M$ estimates, we solve for the relative kinematics $\bYu_M$ (in Section \ref{sec:relativeKinematics}), and the absolute kinematics $\bY_M$ (in Section \ref{sec:absKinematics}), using (\ref{eq:BM_rel_unknown}) and (\ref{eq:BM_abs_unknown}) respectively.
\end{itemize}

Finally, given the initial relative position and the node kinematics, the time-varying absolute and relative positions $\{\bS(t), \bSu(t)\}$  can be estimated using (\ref{eq:absPos}) and (\ref{eq:relPosX2}) respectively.

\begin{figure}[tp] \centering
\scalebox{0.3}[0.3]{\includegraphics{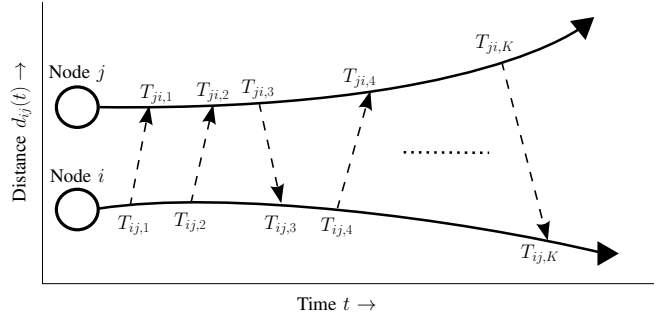}

\rput(-26.5,11.2) {\Huge Node $j$}
\rput(-23,  10.3) {\Huge$T_{ji,1}$}
\rput(-20.5,10.4) {\Huge$T_{ji,2}$}
\rput(-18.5,10.5) {\Huge$T_{ji,3}$}
\rput(-14,  11)   {\Huge$T_{ji,4}$}
\rput(-8,   12.5) {\Huge$T_{ji,K}$}

\rput(-26.5,6.7)  {\Huge Node $i$}
\rput(-24,  4.5)  {\Huge$T_{ij,1}$}
\rput(-21.5,4.6)  {\Huge$T_{ij,2}$}
\rput(-17.5,4.5)  {\Huge$T_{ij,3}$}
\rput(-15,  4.4)  {\Huge$T_{ij,4}$}
\rput(-6,   3.2)  {\Huge$T_{ij,K}$}

\rput(-15,  1.0) {\Huge Time $t$ $\rightarrow$}
\put(-825,250){\rotatebox{90}{\makebox(0,0){\Huge  Distance $d_{ij}(t) \rightarrow$ }}}
}
\caption{ \small \emph{\textbf{Dynamic ranging:}} A Generalized Two Way Ranging (GTWR) scenario between a pair of mobile nodes, where the nodes exchange $K$ time stamps asymmetrically with each other \cite{rajanJ1}. The curved lines symbolize the non-linear motion of the mobile nodes with time. Unlike our previous models  \cite{rajan2012, rajanJ2} which considered only linear independent velocities of the nodes, in this article we consider non-linear motion of the nodes.}
\label{fig:figPairWise}
\end{figure}
 

\section{Dynamic ranging and Relative position} \label{sec:dynamicRanging} In this section, we aim to estimate the range parameters $(\bR, \bdR, \bddR, \hdots)$, given two-way communication between the nodes in the mobile network. In Section \ref{sec:time-varying-tau}, we relate the time-varying propagation delay between the nodes and the range parameters. Given this relationship, we present a Dynamic ranging model in Section \ref{sec:dynamicRangingModel}, and subsequently present a closed form algorithm to estimate the range parameters in Section \ref{sec:dynamicRangingAlgorithm}. Finally, we apply the MDS algorithm to find the initial relative position of the nodes in Section \ref{sec:relativePositionAlgorithm}.

\subsection{Time-varying propagation delay} \label{sec:time-varying-tau}

Consider a pair of mobile nodes capable of communicating with each other. Let $\tau_{ij} (t_0) \triangleq \tau_{ji}(t_0) = c^{-1}d_{ij} (t_0)$ be the propagation delay of this communication between the node pair $(i, j)$ at time instant $t_0$, where $d_{ij}(t_0)$ is the corresponding pairwise distance and $c$ is the speed of the electromagnetic wave in the medium. Now, for a
small interval $\Delta t = t - t_0$, we assume the relative distance to
be a smoothly varying polynomial of time which enables us to describe the propagation delay $\tau_{ij}(t)$ at $t$ as an infinite Taylor series in the neighborhood of $t_0$ \begin{equation}
\tau_{ij}(t)= c^{-1}d_{ij}(t) = 
\underline{r}_{ij} + 
\underline{\dot{r}}_{ij}(t-t_0) + 
\underline{\ddot{r}}_{ij}(t-t_0)^2+ \hdots,
\label{eq:tauij_dij_t_model}
\end{equation} where the Taylor coefficients are defined as \begin{equation}  \label{eq:rangeTranslation} \begin{bmatrix} \underline{r}_{ij}, \underline{\dot{r}}_{ij}, \underline{\ddot{r}}_{ij}, \hdots \end{bmatrix}^T=
\diag(\bgamma)^{-1}\begin{bmatrix} r_{ij}, \dot{r}_{ij}, \ddot{r}_{ij}, \hdots \end{bmatrix}^T,
\end{equation} and $\bgamma= c\ [0!, 1!, 2!, \hdots]^T$. Here, $(r_{ij}, \dot{r}_{ij}, \ddot{r}_{ij}, \hdots)$ are the derivatives of the time-varying pairwise distance $d_{ij}(t)$ esimtated at $t=t_0$, which are the elements of the matrices $(\bR, \bdR, \bddR, \hdots)$, presented earlier in Section \ref{sec:datamodel:distance}. The physical significance of these coefficients is as follows. The pairwise distance at $t_0$ is $r_{ij}$, which is conventionally obtained from time of arrival measurements. $\dot{r}_{ij}$ is the radial velocity, typically observed from Doppler shifts, and the second-order range parameter $\ddot{\br}_{ij}$ is the rate of radial velocity between the node pair at $t_0$. We will now use this relation in a scenario where mobile nodes are capable of two-way communication.

\subsection{Data model} \label{sec:dynamicRangingModel} Consider a Generalized Two Way Ranging (GTWR) scenario between a pair of mobile nodes (\figurename~\ref{fig:figPairWise}), where the nodes communicate asymmetrically with each other, and record $K$ timestamps on each node. The timestamps recorded at the $k$th time instant ($k<K$) at node $i$ and node $j$ are given by $T_{ij,k}$ and $T_{ji,k}$ respectively. The nodes are mobile during these timestamp exchanges, and therefore the propagation delay between the nodes is unique at every time instant. With an abuse of notation, let $\tau_{ij,k}$ and $d_{ij,k}$ be the propagation delay and the distance between the node pair $(i,j)$ at the $k$th time instant. Then assuming the distance is (approx) constant during the propagation time of the message,the non-relativistic propagation delay is $ \tau_{ij,k}= c^{-1}d_{ij,k}=|T_{ij,k} - T_{ji,k}|$. Now, observe that the pairwise propagation delay for GTWR can also be written as (\ref{eq:tauij_dij_t_model}), by replacing $t$ with $T_{ij,k}$ (or $T_{ji,k}$). More concretely, the propagation delay $\tau_{ij}$ is given as \begin{equation}
\tau_{ij,k} 
=|T_{ij,k} - T_{ji,k}|
= \underline{r}_{ij} + 
\underline{\dot{r}}_{ij}(T_{ij,k}-T_0) + 
\underline{\ddot{r}}_{ij}(T_{ij,k}-T_0)^2+ \hdots 
\label{eq:tauij_dij_t_model_k}, 
\end{equation} where the range parameters are estimated at $T_0$ where $T_{ij,k} \le T_0 \le T_{ij,K}$.


Aggregating all the $K$ timestamps for each node pair $(i,j)$, and populating all measurements from $\barN\triangleq 0.5N(N-1)$ unique pairwise links for a network of $N$ nodes, we have \begin{equation} 
\overbrace{\begin{bmatrix} \bI_{\bar{N}} \otimes \b1_K \quad \bT \quad \bT^{\odot 2} \quad \hdots\ \end{bmatrix}}^{\bV} 
\overbrace{\begin{bmatrix} \bur \\ \bdur \\ \bddur  \\ \vdots\ \end{bmatrix}}^{\btheta} 
= \btau, \label{eq:globalNormal} \end{equation} where for an $L$th order polynomial approximation, $\btheta \in \mathbb{R}^{\barN L \times 1}$ is a vector of unknown coefficients. The $\bar{N}$ dimensional vector $\underline{\br}$ = $[\underline{r}_{ij}]\ \forall 1\le i\le N, j\le i $ contains all the pairwise distances at $t_0$, and vectors containing the higher order derivatives $(\underline{\bdr}, \underline{\bddr}, \hdots)$ are similarly defined. The matrix $\bV$ is a Vandermonde-like matrix defined as $\bV =
[\bI_{\bar{N}} \otimes \b1_K \quad \bT \quad \bT^{\odot 2} \quad \hdots\ ]\in \mathbb{R}^{\bar{N}K \times \bar{N}L}$, where $\bT =
\bdiag(\bt_{12}, \bt_{13}, \hdots \bt_{1N},\ \bt_{23}, \hdots )  
\in \mathbb{R}^{\bar{N}K \times \bar{N}}$ and $\bt_{ij}=
[T_{ij,1}-T_0, T_{ij,2}-T_0, \hdots, T_{ij,K}-T_0]^T \in \mathbb{R}^{K \times 1}$ contain all the time stamps. All the unique pairwise propagation delays are collected in $\btau= [\btau^T_{12}, \btau^T_{13}, \hdots \btau^T_{1N},\ \btau^T_{23}, \hdots]^T\in\mathbb{R}^{{N}K \times 1}$ where $\btau_{ij}= | \bt_{ji} - \bt_{ij} |$. Our goal in the following section, is to estimate the values $\begin{bmatrix} \underline{r}_{ij}, \underline{\dot{r}}_{ij}, \underline{\ddot{r}}_{ij}, \hdots \end{bmatrix}$ from (\ref{eq:globalNormal}), which will help us construct the range matrices \big($\bR, \bdR, \bddR, \hdots$\big).

\subsection{Dynamic ranging algorithm} \label{sec:dynamicRangingAlgorithm} 
In reality, the propagation delay is erroneous and hence, more practically (\ref{eq:globalNormal}) is \begin{equation} 
\hat{\btau}= \bV\btheta + \bEta, 
\label{eq:linearVanderModel}
\end{equation} where $\hat{\btau}$ is the noisy propagation delay, and the noise parameters plaguing the data model are populated in $\bEta$= $[\bEta^T_{12}, \bEta^T_{13},$ $\hdots \bEta^T_{1N},\ \bEta^T_{23}, \hdots]^T \in\mathbb{R}^{\bar{N}K \times 1}$, where $\bEta_{ij} =[\eta_{ij,1}, \eta_{ij,2}, \hdots, \eta_{ij,K}]$ is the error unique to the node pair $(i,j)$. In practice, the noise are on the time markers $T_{ij,k}$ and subsequently on the Vandermonde matrix, which has been simplified under nominal assumptions to arrive at the elegant model (\ref{eq:linearVanderModel}). The approximations involved are discussed in Appendix \ref{ap:approxNoiseModel}. 

Now, suppose the covariance of the noise on the normal equations \begin{equation} \label{eq:covNoise}
\bSigma\triangleq\ \mathbb{E}\big\{ \bEta\bEta^T \big\},
\end{equation} is known and invertible, then the weighted least squares solution $\hat{\btheta}$ is obtained by minimizing the following $l_2$ norm, \begin{eqnarray} \label{eq:dyamicWLS}
\hat{\btheta}
&=& \argmin_{\btheta}\norm{\bSigma^{-1/2}(\bV\btheta - \hat{\btau})}^2 \nonumber \\
&=&\big(\bV^T\bSigma^{-1}\bV\big)^{-1}\bV^T\bSigma^{-1}\hat{\btau}.
\end{eqnarray} A valid solution is feasible if $K\ge L$ for each of the $\barN$ pairwise links. More generally, when  $L$ is unknown, an order recursive least squares can be employed to obtain the range coefficients \cite{rajanJ2}. Given $\btheta$, estimates of the range parameter matrices \big($\hat{\bR}, \hat{\bdR}, \hat{\bddR}, \hdots$\big) can be constructed using (\ref{eq:rangeTranslation}) and subsequently, from (\ref{eq:Bt_2}) we have the following estimates
\begin{subequations} \label{eq:Bt_2_estimate} 
\begin{eqnarray} 
\hat{\bBu}^{(0)}       
&=& -0.5\bP\hat{\bR}^{\odot 2}\bP, 
\label{eq:Bt_2a_estimate}                    \\
\hat{\bBu}^{(1)}
&=& -\bP\Big[ \hat{\bR}\odot \hat{\bdR} \Big]\bP, 
\label{eq:Bt_2b_estimate}            \\
\hat{\bBu}^{(2)}
&=& -\bP \Big[ \hat{\bR}\odot\hat{\bddR} + \hat{\bdR}^{\odot 2} \Big]\bP. 
\label{eq:Bt_2c_estimate} \end{eqnarray} \end{subequations}
 
\color{black}

\subsection{Relative position} \label{sec:relativePositionAlgorithm} Give the initial pairwise distances at $t_0$ \ie $\bR$, the initial relative positions $\bXu$ can be determined via MDS. Given $\hat{\bR}$, let $\hat{\bB}_0$ be an estimate of $\bB_0\triangleq \bBu^{(0)}$, obtained using (\ref{eq:Bt_2a_estimate}). A spectral decomposition of this matrix yields $\hat{\bB}_0= \bV_x\bLambda_x\bV^T_x$, where $\bLambda_x$ is an $N$ dimensional diagonal matrix containing the eigenvalues of the $\hat{\bB}_0$ and $\bV_x$ the corresponding eigenvectors. An estimate of the relative position estimate using MDS is then given by \begin{eqnarray}
\hat{\bXu}
&=&\argmin_{\bXu}\ \norm{\hat{\bB}_0- \bXu^T\bXu}\ \text{s.t.}\ \text{rank}(\bXu)= P \nonumber \\
&=&\underline{\bLambda}^{1/2}_x\underline{\bV}^T_x \label{eq:posMDS},
\end{eqnarray}  where $\underline{\bLambda}_x$ contains the first $P$ nonzero eigenvalues from $\bLambda_x$ and $\underline{\bV}_x$ is a subset of $\bV_x$ containing the corresponding eigenvectors \cite{borg97}.

\section{Relative kinematics} \label{sec:relativeKinematics} 

In the previous section, we estimated the range parameters given time-varying distance measurements $\bD(t)$, which was the first step (S1) in our problem statement described in Section \ref{sec:datamodel:problemStatement}. Using these range parameters, we constructed the double centered matrices $\big(\hat{\bBu}^{(0)}, \hat{\bBu}^{(1)}, \hat{\bBu}^{(2)}, \hdots \big)$ (\ref{eq:Bt_2_estimate}) and estimated the relative position $\hat{\bXu}$  using MDS (\ref{eq:posMDS}). Given these estimates, we now aim to solve the unknown relative kinematic matrices $\bYu_M$ using (\ref{eq:BM}), as proposed in (S2) of Section \ref{sec:datamodel:problemStatement}.
\color{black}

\subsection{Linearized multidimensional scaling (LMDS)} Prior to investigating the general kinematic model (\ref{eq:BM}), we revisit a special case when the nodes are mobile under linear independent motion \cite{rajanJ2}. In such a scenario, the acceleration and other higher order derivatives are absent \ie $\bY_m = \bzero,\ \forall\ m \ge 2$. Therefore, under constant velocity assumption, equations (\ref{eq:Bt_3b}) and (\ref{eq:Bt_3c}) simplify to \begin{subequations}\begin{eqnarray}
\bBu^{(1)}&=&\ \bXu^T\bYu_1 + \bYu_1^T\bXu,	
\label{eq:lmds1_1} \\
\bBu^{(2)}&=&\ 2\bYu_1^T\bYu_1 
\label{eq:lmds2_1},
\end{eqnarray} \end{subequations} and for $m \ge 3$ $\{\bB_m, \bBu^{(m)} \}$ defined in (\ref{eq:BM}) does not exist \cite[Appendix B]{rajanJ2}. Now substituting the definition of relative velocity from (\ref{eq:relToAbsMthorder}) and exploiting the property $\bH^T_1\bH_1 = \bI$, we have  \begin{subequations}\begin{eqnarray}
\bBu^{(1)} &=&\ \bXu^T\bH_1\tilde{\bYu}_1 + \tilde{\bYu}_1^T\bH^T_1\bXu,	\label{eq:lmds1} \\
\bBu^{(2)} &=&\ 2\tilde{\bYu}_1^T\tilde{\bYu}_1 \label{eq:lmds2}.
\end{eqnarray} \end{subequations} The LMDS algorithm to estimate the relative velocity (upto a translation) is then a two step method as decribed below.

\subsubsection{MDS-based relative velocity estimator} Firstly, the relative velocity upto a rotation and translation is obtained by minimizing the strain function using (\ref{eq:lmds2}). Let $\hat{\bBu}^{(2)}$ be an estimate of $\bBu^{(2)}$ from (\ref{eq:Bt_2c_estimate}), with an eigenvalue decomposition $\hat{\bBu}^{(2)} \triangleq \bV_y\bLambda_y \bV^T_y$, then the relative velocity estimate is given by \begin{eqnarray}
\hat{\tilde{\bYu}}_1
&=&\argmin_{\tilde{\bYu}_1}\ 
\norm{\hat{\bBu}^{(2)}- 2\tilde{\bYu}^T_1\tilde{\bYu}_1}\ 
\text{s.t.}\ \text{rank}(\tilde{\bYu}_1)= P \nonumber \\
&=&\underline{\bLambda}^{1/2}_y\underline{\bV}^T_y \label{eq:velMDS}, 
\end{eqnarray} where $\underline{\bLambda}_y$ and $\underline{\bV}_y$ contain the first $P$ nonzero eigenvalues and corresponding eigenvectors of $\bLambda_y$ and $\bV_y$ respectively.

\subsubsection{Estimating the unknown rotation} The MDS-based solution (\ref{eq:velMDS}) yields the relative velocity upto a rotation and translation, which is not sufficient to reconstruct the time-varying relative position using (\ref{eq:relPosX1}). To estimate the unique rotation matrix, we vectorize (\ref{eq:lmds1}), apply the transformation (\ref{eq:vecABC}), and solve the following constrained cost function \begin{equation}
\argmin_{\bH_1}\ \norm{\hat{\bPhi}\text{vec}(\bH_1)- \text{vec}{(\hat{\bBu}^{(1)})}}^2
\;\;\text{s.t}\quad \bH_1^T\bH_1= \bI_P,
\label{eq:H1_cls} \end{equation} where $\hat{\bPhi}= (\bI_{N^2} + \bJ)(\hat{\tilde{\bYu}}_1^T \otimes \hat{\bXu}^T)$, $\{\hat{\bXu},\hat{\tilde{\bYu}}_1\}$ are estimates obtained from (\ref{eq:posMDS}) and (\ref{eq:velMDS}) respectively and, $\bJ$ is a permutation matrix such that (\ref{eq:vecA}) holds.

Thus, under linear velocity assumption, the relative velocity $\bYu_1= \bH_1\tilde{\bYu}_1$ up to a translation can be reconstructed for a general $P$ dimensional scenario using the estimators (\ref{eq:velMDS}) and (\ref{eq:H1_cls}). It is worth noting that the LMDS solution is feasible, only under the constant velocity assumption. In general, the assumption on linear motion is not always valid and hence we address the more general kinematic motion in the following sections.

\subsection{Lyapunov-like equations} More generally, when the nodes are in non-linear motion, the kinematics $\bY_m\ \forall\ m \ge 1$ exists and must be estimated. To solve for the relative kinematics in this scenario, we refer back to our relative kinematic model (\ref{eq:BM}). For any $M \ge 1$, the model (\ref{eq:BM_rel_unknown}) 
\begin{equation}\label{eq:lyapunovLike}
\bB_{M} = \bXu^T\bYu_{M} + \bYu^T_{M}\bXu,
\end{equation} is the relative \emph{Lyapunov-like equation} \cite{hodges1957,chiang2012}, where $\bB_M$ is the $N-$dimensional measurement matrix and $\bY_M$ is the $M$th order kinematics to be estimated. As pointed out in Remark 1 in Section \ref{sec:datamodel:datamodel}, $\bB_M$ can be constructed by $\bB^{(M)}$ and lower order relative kinematics $\{\bY_m\}^{M-1}_{m=1}$. The above equation is very similar, but not the same as the following equations, \begin{eqnarray}
\bA^H\bY + \bY\bA 	&=& \bB, \nonumber \\
\bA\bY + \bY\bA			&=& \bzero, \nonumber \\
\bA\bY + \bY\bC 	 	&=& \bE, \nonumber
\end{eqnarray} which are the (continuous)  \emph{Lyapunov equation}, \emph{commutativity equation} \cite[chapter 4]{roger1994topics} and \emph{Sylvester equation} \cite{bartels1972,golub1979} respectively, where the unknown matrix $\bY$ has to be estimated, given $\bA, \bB, \bC, \bE$. The solutions to these equations exist and are extensively investigated in control theory literature \cite{bhatia1997}. However the Lyapunov-like equation (\ref{eq:lyapunovLike}) has received relatively less attention. The Lyapunov-like equation has a straight forward solution for $P=1$. But, for $P\ge 2$, although a general solution was proposed by Braden \cite{braden1998}, a unique solution to (\ref{eq:lyapunovLike}) does not exist which we discuss in \appendixname\ \ref{ap:underdeterminedLyapunov}.

Now, vectorizing (\ref{eq:lyapunovLike}) and using (\ref{eq:vecABC}), we aim to solve \begin{eqnarray} 
\label{eq:vecRelLyapunov} 
\hat{\byu}_M 	
&=& \argmin_{\byu_M} \norm{(\bI_{N^2} + \bJ)(\bI_N \otimes \bXu^T)\byu_M  - \bb_M}^2 \nonumber \\
&=& \argmin_{\byu_M}	\norm{\bAu\byu_M - \bb_M}^2 ,
\label{eq:costFunctionRel}
\end{eqnarray} where \begin{subequations} \begin{eqnarray}
\bAu  &=& (\bI_{N^2} + \bJ)(\bI_N \otimes \bXu^T)\in \mathbb{R}^{N^2 \times NP}, \label{eq:ADef}\\
\byu_M&=& \vect(\bYu_M) \in \mathbb{R}^{NP \times 1}, \label{eq:yDef}\\
\bb_M &=& \vect(\bB_M) 	\in \mathbb{R}^{NP \times 1}, \label{eq:bDef} \end{eqnarray} 
\end{subequations} and $\bJ$ is an orthogonal permutation matrix (\ref{eq:vecA}). The matrix $(\bI_N \otimes \bXu^T) \in \mathbb{R}^{N^2 \times NP}$ is full column rank, since $\bX$ is typically non-singular. However, the sum of permutation matrices $(\bI_{N^2} + \bJ)\in \mathbb{R}^{N^2 \times N^2}$ is always rank deficient by at least ${N \choose 2}$. Hence, the matrix primary objective function $\bAu$ is not full column rank, but is rank deficient by at least $\barP \triangleq\ 0.5P(P-1)$, which is discussed in Appendix \ref{ap:underdeterminedLyapunov}. In (\ref{eq:lyapunovLike}),  since the translational vectors of both $\bX$ and $\bY_M$ are projected out using the centering matrix $\bP$, the $\barP$ dependent columns in $\bAu$ indicate the rotational degrees of freedom in a $P$-dimensional Euclidean space.

\subsection{Lyapunov-like least squares (LLS)} A unique solution to the Lyapunov-like equation is not feasible without sufficient constraints on the linear system (\ref{eq:vecRelLyapunov}). Let $\hat{\bA}$ be an estimate of the $\bA$, obtained by substituting the estimated relative position $\hat{\bXu}$ (\ref{eq:posMDS}). Similarly, let $\hat{\bb}_M$ be an estimate of $\bb_M$ obtained by substituting the range parameters and appropriate relative kinematic matrices upto order $M-1$. Then the constrained Lyapunov-like least squares (LLS) solution to estimating the relative kinematic matrices is given by minimizing the cost function \begin{equation}
\hat{\byu}_{M,lls} 
= \argmin_{\byu_M}\norm{\hat{\bAu}\byu_M - \hat{\bb}_M}^2	
\quad \text{s.t.}\ \bar{\bC}\byu_M=\bar{\bd}, \label{eq:cclsRel}
\end{equation} where $\bar{\bC}$ is a set of non-redundant constraints. The above optimization problem has a closed-form solution, given by solving the KKT equations (\appendixname\ \ref{ap:KKT}).


\subsection{Weighted Lyapunov-like LS (WLLS)} In reality, both $\bA$ and $\bb$ are plagued with errors and hence the solution to the cost function (\ref{eq:cclsRel}) is sub-optimal. Let $\bar{\bW}$ be an appropriate weighting matrix on the Lyapunov-like equation, then the Weighted Lyapunov-like Least Squares (WLLS) solution is obtained by minimizing the cost function \begin{equation}
\hat{\byu}_{M,wlls}
=\argmin_{\byu_M}\ \norm{\bar{\bW}_M^{1/2}\big( \hat{\bAu}\byu_M- \hat{\bb}_M \big)}^2 
\quad \text{s.t.}\ \bar{\bC}\byu_M=\bar{\bd},  \label{eq:wclsRel} \\
\end{equation} which, similar to (\ref{eq:cclsRel}), can be solved using the constrained KKT solutions (\appendixname\ \ref{ap:KKT}). An appropriate choice of the weighting matrix $\bar{\bW}_M$ will be discussed in Section \ref{sec:crb_W}.
\begin{flushleft}

\end{flushleft}

\subsection{Choice of constraints: Relative immobility} In the absence of absolute location information, a unique solution is feasible if the relative motion of at least $P$ nodes or features are invariant (or known) over a small time duration $\Delta t$. In an anchorless framework, a set of given nodes would have equivalent relative kinematics, if they are identical in motion upto a translation or if they are immobile for the small measurement time $\Delta t$. Such situations could arise, for example, in underwater localization, when a few immobile nodes could be fixed with unknown absolute locations, which in turn could assist the relative localization of the other nodes. For $P=2$, if the first $P$ nodes are relatively immobile for the small measurement time, a valid constraint for (\ref{eq:cclsRel}) and (\ref{eq:wclsRel}) is \begin{equation} \label{eq:constraintImmobile}
\bar{\bC}_1=	\begin{bmatrix} \bI_2& -\bI_2& \bzero \end{bmatrix}, 
\qquad
\bar{\bd}_1=	\bzero,
\end{equation} which can be readily extended for $P>2$ and if required, for a larger number of immobile nodes. In essence, the relative immobility constraint reduces the parameter space in pursuit of a unique solution for the ill-posed Lyapunov-like equation.

\subsection{Time-varying relative position} In this section, we solved for the relative kinematics of motion, using the range parameters and relative position estimates. When the nodes are in linear motion, the first-order relative kinematics can be estimated using the LMDS algorithm (\ref{eq:velMDS}, \ref{eq:H1_cls}). More generally, for estimating the relative kinematics in a non-linear scenario, we solve the Lyapunov-like equation (\ref{eq:lyapunovLike}) using constrained Least squares (\ref{eq:cclsRel}, \ref{eq:wclsRel}). Substituting these estimates in (\ref{eq:relPosX2}), an estimate of the relative time-varying position is \begin{eqnarray} \label{eq:relPos_estimate} \hat{\underline{\bS}}(t)&=&
\hat{\bXu} + \hat{\bYu}_1(t-t_0) 	+ 0.5\hat{\bYu}_2 (t-t_0)^2+ \hdots 
\end{eqnarray} where $\hat{\bXu}$ is a relative position estimate from (\ref{eq:posMDS}) and $\{\hat{\bYu}_1, \hat{\bYu}_2, \hdots \}$ are the estimates from (\ref{eq:cclsRel}) or (\ref{eq:wclsRel}). In the following section, we aim to estimate the absolute kinematics of the nodes and subsequently the time-varying absolute position. 

%
%

\section{Absolute Kinematics} \label{sec:absKinematics}

%

In this section, we solve for the absolute kinematics $\bY_M$, given $\bB_M$ and the relative position $\bXu$. We have from (\ref{eq:BM_abs_unknown}),  \begin{equation} \label{eq:absoluteLyapunov}
\bXu^T\bY_M\bP + \bP\bY_M^T\bXu = \bB_M.  
\end{equation} The above equation is similar, but not the same, to the generalized (continuous-time) Lyapunov equation \begin{equation}
\bA^T\bY\bC + \bC^T\bY\bA = \bB, \nonumber
\end{equation} where $\bA, \bB, \bC$ are known square matrices \cite{penzl1998numerical}. We now vectorize (\ref{eq:absoluteLyapunov}) and aim to minimize the following cost function \begin{equation} 
\hat{\by}_M=\ \argmin_{\by_M} \norm{\bA\by_M - \bb_M }^2,
\label{eq:costFunctionAbs}
\end{equation} where \begin{subequations} \begin{eqnarray}
\bA   &=& (\bI_{N^2} + 	\bJ)(\bP \otimes \bXu^T)\in \mathbb{R}^{N^2 \times NP}, \label{eq:ADefabs}\\
\by_M	&=& \vect(\bY_M) \in \mathbb{R}^{NP \times 1}, \label{eq:yDefabs} \end{eqnarray} \end{subequations} and $\bb_M$ is given by (\ref{eq:bDef}). In comparison to (\ref{eq:vecRelLyapunov}), the matrix $(\bI_{N} \otimes \bXu^T)$  is replaced with $(\bP \otimes \bXu^T)$ in (\ref{eq:ADefabs}). The rank of the centering matrix $P$ is $N-1$ and since $\bXu$ is typically full row rank, the Kronecker product is utmost of rank $NP-P$. This rank-deficiency of $P$ is also reflected in the matrix $\bA$. Unlike $\bAu$ which has $\barP$ dependent colomns, $\bA$ is rank-deficient by ${P+1 \choose 2}= \barP + P$. The additional $P$ dependent columns are perhaps not surprising, as they indicate the lack of information on the translational vector, \ie the group center of the $M$th order kinematic matrix. 

\subsection{Generalized Lyapunov-like least squares (GLLS)} In pursuit of a unique solution to the rank-deficient system (\ref{eq:costFunctionAbs}), we propose a constrained generalized Lyapunov-like least squares (GLLS) to estimate the absolute kinematic matrices which is obtained by minimizing the cost function \begin{equation} \hat{\by}_{M,glls} = \argmin_{\by_M}\norm{\hat{\bA}\by_M - \hat{\bb}_M}^2	
\quad \text{s.t.}\ \bC\by_M=\bd, \label{eq:cclsAbs}
\end{equation} where $\hat{\bA}$ and $\hat{\bb}_M$ are estimates of $\bA$ and $\bb_M$ respectively. The matrix $\bC$ is a set of non-redundant constraints, which will be discussed in Section \ref{sec:constraintsLyapunov_abs}.

\subsection{Weighted generalized Lyapunov-like LS (WGLLS)} The performance of the estimator can be improved by weighting the cost function (\ref{eq:cclsAbs}), \ie \begin{equation} \hat{\by}_{M,wglls}=\argmin_{\by_M}\norm{\bW_M^{1/2}\big( \hat{\bA}\by_M- \hat{\bb}_M \big)}^2  \quad \text{s.t.}\ \bC\by_M=\bd,  \label{eq:wclsAbs} \\ \end{equation} which yields the weighted generalized Lyapunov-like least squares (WGLLS) solution (see \appendixname\ \ref{ap:KKT}), where $\bW_M$ is an appropriate weighting matrix (see Section \ref{sec:crb_W}).

\subsection{Choice of constraints: Anchor-aware network}\label{sec:constraintsLyapunov_abs} For an anchored scenario, if the $M$th order absolute kinematics of a few nodes are known, then the absolute velocity, acceleration and higher order derivatives can be estimated. A straightforward minimal constraint for the feasible solution is then \begin{equation}\label{eq:constraintKnown}
\bC_1 = \begin{bmatrix} \bI_{\bar{P}+ P},& \bzero \end{bmatrix}, 
\end{equation} where without loss of generality, we assume the first $\bar{P}+P$ parameters are known. 


\subsection{Time-varying absolute position} In (\ref{eq:cclsAbs}, \ref{eq:wclsAbs}), we solved for the absolute kinematics given the measurement matrix $\bB_M$ and the relative position, using constrained Least squares estimators. Given these estimates, we have from (\ref{eq:absPos}) \begin{eqnarray}\label{eq:absPos_estimate}
\hat{\bS}(t) &=& 
\hat{\bXu} + \hat{\bY}_1(t-t_0) 	+ 0.5\hat{\bY}_2 (t-t_0)^2+ \hdots,
\end{eqnarray} where $\hat{\bS}(t)$ is an estimate of the time-varying absolute position, $\hat{\bXu}$ is an estimate of the relative position (\ref{eq:posMDS}), and $\{\hat{\bY}_1, \hat{\bY}_2, \hdots \}$ are the absolute kinematic estimates obtained by solving (\ref{eq:cclsAbs}) or (\ref{eq:wclsAbs}).

\section{\Cramer-Rao Bounds} \label{sec:crb}  The \Cramer-Rao lower Bound (CRB) sets a lower bound on the minimum achievable variance of any unbiased estimator. In this section, we derive the CRBs for the estimated parameters based on the presented data models. In the following section, we will use these bounds to benchmark the performance of the proposed estimators.

\subsection{Range parameters} \label{sec:crb_range} We begin by deriving the lower bounds on the range parameters. Let $\bpsi=[\br^T, \bdr^T, \bddr^T, \hdots]^T$ and let $\hat{\bpsi}$ be the corresponding estimate, then the CRB $\bSigma_{\psi}
\triangleq\
{\mathbb{E}} 
\left \{ 
(\hat{\bpsi}-\bpsi)(\hat{\bpsi}-\bpsi)^T 
\right \} $ on the range parameters for the linear model ($\ref{eq:linearVanderModel}$) is \begin{equation} 
\label{eq:crbTheta}
\bSigma_{\psi}
\ge 
\bGamma\big(\bV^T\bSigma^{-1}\bV\big)^{-1}\bGamma
= \begin{bmatrix}
\bSigma_r & * & * & *  \\
*					& \bSigma_{\dr} & * & * \\
* & * & 		\bSigma_{\ddr} & * \\
* & * & 	*& \ddots 
\end{bmatrix},
\end{equation} where $\bSigma_{\psi}$ is the covariance of $\bpsi$ and  $\bSigma$ is the covariance of the noise on the timestamps defined in (\ref{eq:covNoise}). Here, the covariance matrices $\{\bSigma_r, \bSigma_{\dr}, \bSigma_{\ddr}, \hdots \}$ are the lowest achievable bounds for the corresponding range parameters $\{\br, \bdr, \bddr, \hdots \}$. The entries not of interest are denoted by $*$ and $\bGamma= \diag(\bgamma) \otimes \bI_{\barN}$ is a transformation matrix, where $\bgamma$ is given by (\ref{eq:rangeTranslation}).  It is worth noting that our proposed solution (\ref{eq:dyamicWLS}) achieves this lower bound for an appropriate $L$.

\color{black}

\subsection{Relative position} \label{sec:crb_pos} The CRB on the relative positions $\by_0 \triangleq\ \vect(\bXu)$ is given by the inverse of the Fisher Information Matrix (FIM) \ie \begin{eqnarray}
\bSigma_x\triangleq {\mathbb{E}} \left \{ (\hat{\by}_0-\by_0)(\hat{\by}_0-\by_0)^T \right \}
\ge \bF_x^{\dagger}, \label{eq:CRBx} 	
\end{eqnarray} where $\hat{\by}_0$ is an estimate of the unknown relative position $\by_0$, $\bSigma_x$ is the covariance of $\hat{\by}_0$ \cite{rajanJ2} and the FIM $\bF_x \in \mathbb{R}^{NP \times NP}$ is \begin{equation}
\bF_x= \bJ_x^T \bar{\bSigma}_r^{-1} \bJ_x,
\label{eq:FIMx}
\end{equation} where $\bar{\bSigma}_r\triangleq\ \bdiag(\bSigma_r, \bSigma_r)$, $\bJ_x$ is the Jacobian \cite[\appendixname\ C]{rajanJ2} and $\bSigma_r$ is obtained from (\ref{eq:crbTheta}). In the absence of known anchors in the network, the FIM  is inherently nonlinear and hence we employ the Moore-Penrose pseudoinverse in (\ref{eq:CRBx}). 

\subsection{Kinematics} We now derive the lower bounds on the variance of the estimates of the relative kinematics $\byu_M= \vect(\bYu_M)$ and absolute kinematics $\by_M= \vect(\bY_M)$. The Gaussian noise vectors plaguing the cost functions (\ref{eq:costFunctionRel}) and (\ref{eq:costFunctionAbs}) are modeled as \begin{eqnarray}
\brhou_M &\sim &
\cN(\bAu\byu_M-\bb_M, \bSigmau_{\rho,M}), \\
\brho_M &\sim& 
\cN(\bA\byu_M-\bb_M, \bSigma_{\rho,M}),
\end{eqnarray} where $\brho_{M}, \brhou_{M}$ are $N^2$ dimensional noise vectors, and the corresponding covariance matrices are of the form \begin{subequations} \label{eq:rhoCov} \begin{eqnarray} 
\bSigmau_{\rho,M}	\triangleq& 
\mathbb{E} \big\{ \brhou_M\brhou_M^T \big\} \approx\
\bAu_{y,M}\bar{\bSigma}_x\bAu_{y,M}^T + \bSigma_{b,M}, \label{eq:rhoCovAbs} \\
\bSigma_{\rho,M}	\triangleq& 
\mathbb{E} \big\{ \brho_M\brho_M^T \big\} \approx\ 
\bA_{y,M}\bar{\bSigma}_x\bA_{y,M}^T + \bSigma_{b,M}, \label{eq:rhoCovRel}
\end{eqnarray} \end{subequations} where \begin{subequations}  \label{eq:unconstrainedA}
\begin{eqnarray} \label{eq:unconstrainedAy}
\bAu_{y,M} 		  	&=& (\bI_{N^2} + \bJ)(\bI_N \otimes \bYu_M^T)\in \mathbb{R}^{N^2 \times NP}, 	\label{eq:Aydef_} \\
\bA_{y,M} 		  	&=& (\bI_{N^2} + \bJ)(\bP 	\otimes \bY_M^T)\in \mathbb{R}^{N^2 \times NP},	\label{eq:Aydef}
\end{eqnarray} \end{subequations} and an expression for $\bSigma_{b,M}$ is derived in \appendixname\ \ref{ap:covariance_B}.

\subsubsection{Unconstrained CRBs} \label{sec:crb_Lyapunov}
The lowest achievable variance by an unbiased estimator is given by \begin{subequations} 
\label{eq:unconstrainedCRB}
\begin{eqnarray}
\bSigmau_{y,M} \triangleq& 	 
{\mathbb{E}} \left \{ (\hat{\byu}_M-\byu_M)(\hat{\byu}_M-\byu_M)^T \right \} \ge\bF_{y,M}^{\dagger}, \\
\bSigma_{y,M} \triangleq& 	{\mathbb{E}} \left \{ (\hat{\by}_M-\by_M)(\hat{\by}_M-\by_M)^T \right \}  \ge \bF_{y,M}^{\dagger}, \end{eqnarray} \end{subequations} where the corresponding FIMs are given by \begin{subequations}  \label{eq:unconstrainedFIM}
\begin{eqnarray} 
\bFu_{y,M}				
&=&	\bAu^T\bSigma_{\rho,M}^{\dagger}\bAu, \\
\bF_{y,M}					
&=&	\bA^T\bSigma_{\rho,M}^{\dagger}\bA.
\end{eqnarray} \end{subequations} It is worth noting that  the Moore-Penrose pseudoinverse is employed since the FIM is rank-deficient, and consequently the derived bounds (\ref{eq:unconstrainedCRB}) are oracle-bounds.

\subsubsection{Constrained CRBs} 
When the FIM is rank-deficient, a constrained CRB can be derived given differentiable and deterministic constraints on the parameters \cite{stoica1998}. Let $\bar{\bU}, \bU$ be an orthonormal basis for the null space of the constraint matrices $\bar{\bC}, \bC$, then the constrained \Cramer-Rao bound (CCRB) on the $M$th order kinematics are given by \begin{subequations}  \label{eq:crbLyapunov} \begin{eqnarray}  \bSigmau^{C}_{y,M}
&\triangleq& 	{\mathbb{E}} \left \{ (\hat{\byu}_M-\byu_M)(\hat{\byu}_M-\byu_M)^T \right \}   \nonumber \\
&\ge& 				\bar{\bU}\big(\bar{\bU}^T\bFu_{y,M}\bar{\bU}\big)^{-1}\bar{\bU}^T, \\
\bSigma^{C}_{y,M}
&\triangleq& 	{\mathbb{E}} \left \{ (\hat{\by}_M-\by_M)(\hat{\by}_M-\by_M)^T \right \}   \nonumber \\
&\ge& 				\bU\big(\bU^T\bF_{y,M}\bU\big)^{-1}\bU^T, \end{eqnarray} \end{subequations} where the FIMs are given by (\ref{eq:unconstrainedFIM}).

\subsection{Choice of weighting matrices $\bar{\bW}_M, \bW_M$} \label{sec:crb_W} To admit a BLUE solution, we use the inverse of the covariance matrices $\bSigmau_{\rho,M}, \bSigma_{\rho,M}$ as weights to solve the regression problems (\ref{eq:wclsRel}) and (\ref{eq:wclsAbs}), \ie \begin{subequations}
\label{eq:rhoCovEstimate}
\begin{eqnarray}
\bar{\bW}_M 	\triangleq\ 
\hat{\bSigmau}_{\rho,M}^{\dagger}= 	
\big(\hat{\bAu}_y\hat{\bar{\bSigma}}_x\hat{\bAu}_y^T + \hat{\bSigma}_{b,M}\big)^{\dagger}, 	\label{eq:rhoCovEstimateRel} \\
\bW_M	\triangleq\ 
\hat{\bSigma}_{\rho,M}^{\dagger}= 	
\big(\hat{\bA}_y\hat{\bar{\bSigma}}_x\hat{\bA}_y^T + \hat{\bSigma}_{b,M}\big)^{\dagger}, 		\label{eq:rhoCovEstimateAbs}
\end{eqnarray} \end{subequations} where the estimates $\hat{\bAu}_y, \hat{\bA}_y$ are obtained by substituting $\hat{\bYu}_M$ from LLS [(\ref{eq:cclsRel}) and (\ref{eq:cclsAbs})], in (\ref{eq:unconstrainedA}), $\hat{\bar{\bSigma}}_x$ is an estimate of (\ref{eq:CRBx}) and $\hat{\bSigma}_{b,M}$ is derived in \appendixname\ \ref{ap:covariance_B} from appropriate range parameter estimates.

\begin{figure*}
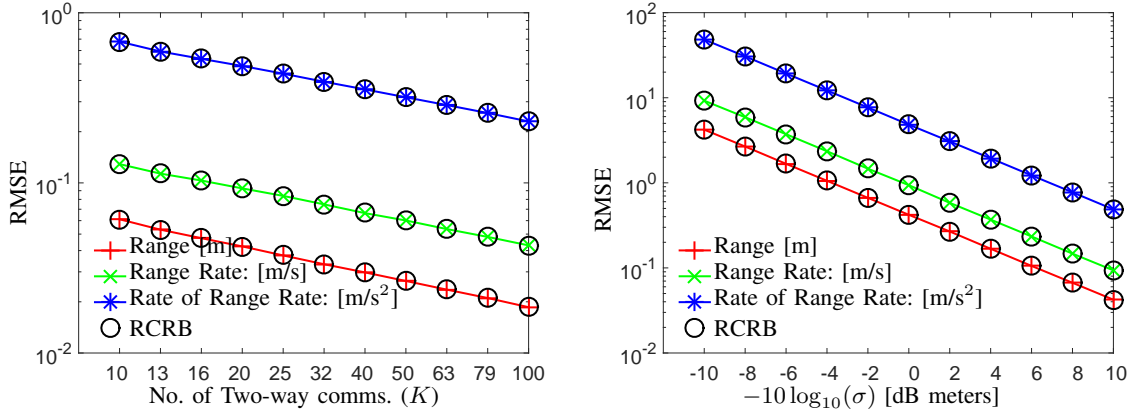
 \centering 
\centering
{   
    \subfloat{
		\psfrag{rT}[cb]	{\small }
		\psfrag{rX}[cb]	{\small No. of Two-way comms. ($K$)}
		\psfrag{rY}[cb]	{\small RMSE}
		\psfrag{rLa}[lb]{\small Range [m]}
		\psfrag{rLb}[lb]{\small Range Rate: [m/s]}
		\psfrag{rLc}[lb]{\small Rate of Range Rate: [m/s$^2$]}
		\psfrag{rLd}[lb]{\small RCRB}	
        \includegraphics[scale=0.38]{range.eps}}
}
\;
{
    \subfloat{
		\psfrag{rST}[cb]{\small  }
		\psfrag{rSX}[cb]{\small  $-10\log_{10}(\sigma)$ [dB meters]}
		\psfrag{rSY}[cb]{\small  RMSE}
		\psfrag{rSLa}[lb]{\small Range [m]}
		\psfrag{rSLb}[lb]{\small Range Rate: [m/s]}
		\psfrag{rSLc}[lb]{\small Rate of Range Rate: [m/s$^2$]}
		\psfrag{rSLd}[lb]{\small RCRB}		
    \includegraphics[scale=0.38]{rangeS.eps}}
}
	\caption{\small \emph{\textbf{Range parameters:}} \textsl{\textbf{Varying $K$ :}} RMSEs (and RCRBs) of relative range parameters $(\br,\bdr, \bddr)$ for varying number of communications ($K$) between the $N=10$ \emph{mobile} nodes for $\sigma=0.1$ meters. \textsl{\textbf{Varying $\sigma$:}} RMSEs (and RCRBs) of relative range parameters $(\br,\bdr, \bddr)$ for a network of $N=10$ nodes exchanging $K=10$ timestamps, where the noise on the time markers ($\sigma$) is varied. Unlike our previous experiments \cite{rajanJ1,rajanJ2}, we consider acceleration in the current setup.}
	\label{fig:rangeSim}
\end{figure*}

\begin{figure*} [tb] \normalsize
\begin{subequations} \label{eq:posVelAccValues}
\begin{eqnarray}
\label{eq:posValues}
\bX&=&
  \begin{bmatrix}
  -244&   385&    81&   -19&  -792&  -554&  -965&  -985&   -49&  -503 \\
  -588&  -456&  -992&  -730&   879&   970&   155&   318&  -858&   419	
  \end{bmatrix} \text{m} \\
\label{eq:velValues}
\bY_1&=&
  \begin{bmatrix}
-5&    -5&    -6&     6&    -1&     2&     1&    -5&     9&    -5 \\
-8&    -8&    -7&    -9&    -3&    -2&    -2&   -10&     2&    -1
  \end{bmatrix} \text{ms}^{-1} \\
\label{eq:accValues}
\bY_2&=&
  \begin{bmatrix}
   -0.17&   -0.17&    0.22&   -0.07&    0.21&   -0.15&    0.55&   -0.72&   -0.49&   -0.34 \\
    0.42&    0.42&    0.98&    0.73&    0.48&    0.08&   -0.43&   -0.14&    0.56&    0.91
  \end{bmatrix} \text{ms}^{-2}
\end{eqnarray}
\end{subequations}
\hrulefill
\end{figure*}

\section{Simulations} \label{sec:simulations} In this section, we conduct experiments to validate the proposed data model, and the solutions against their respective derived lower bounds. A network of $N=10$ nodes is considered in $P=2$ dimensional space, with instantaneous position, velocity and acceleration values arbitrarily chosen as in (\ref{eq:posVelAccValues}), such that the constraint (\ref{eq:constraintImmobile}) holds. All the nodes communicate with each other within a small time-interval of $\Delta T= [T_{ij,k}, T_{ji,k}]=[-1,1]$ seconds, wherein the transmit time markers are chosen to be linearly spaced Without loss of generality, we are interested in the instantaneous kinematics of the nodes at time instant $t_0=T_0=0$. \color{black}

We assume that  all the pairwise communications are independent of each other, \ie $\bSigma= \sigma^2\bI_{\bar{N}K}$. The metric used to evaluate the performance of the range parameters is the root mean square error (RMSE), given by 
\begin{equation} \label{eq:rmseAbs}
\text{RMSE}(\bz)= N_z^{-1} \sqrt{N^{-1}_{exp} \sum^{N_{exp}}_{n=1}\norm{\hat{\bz}(i)-\bz}^2}, 
\end{equation} where $\hat{\bz}(i)$ is the estimate of the unknown vector $\bz \in \mathbb{R}^{N_z \times 1}$ related to the $i$th run of $N_{exp}=500$ Monte Carlo runs. To evaluate the estimates of the relative and absolute kinematic matrices, we use $\bz= \text{vec}(\bU)$, where $\bU$ is the matrix under evaluation. To qualify these estimates, the square root of the \Cramer-Rao Bound (RCRB) is plotted along with the respective RMSE. It is worth noting that the theoretical lower bounds for the range parameters (\ref{eq:crbTheta}), and subsequently the bounds for relative position (\ref{eq:FIMx}) and node kinematics (\ref{eq:unconstrainedCRB}, \ref{eq:crbLyapunov}) are dependent on the covariance of the noise on time markers \ie $\bSigma$. 

For all the proposed estimators in Sections \ref{sec:simulations}\ A-C , we conduct two types of experiments. Firstly, for (a) varying number of pairwise communications $K$ from $0$ to $100$, with constant noise of $\sigma=0.1$m, and secondly for (b) varying SNR from $[-10,10]$ dB meter with a fixed $K=10$ time-stamp exchanges. The noise considered on the time-markers is typical of TWR based fixed localization experiments \cite{patwari2003}.

\subsection{Range parameters} We employ the dynamic ranging algorithm (\ref{eq:dyamicWLS}) for $L=3$, to estimate the desired range coefficients from the time-varying propagation delays. In comparison to our previous experiments \cite{rajanJ1, rajanJ2}, we additionally consider acceleration in the current simulation. \figurename~\ref{fig:rangeSim} shows the RMSE and RCRB of the first $3$ range coefficients, for both varying $K$  and varying SNR, where we observe that the RMSEs achieve the corresponding derived RCRBs asymptotically. Observe that in the Monte carlo experiments, we consider the noise on the time makers, whereas the lower bounds are derived on the data model with approximated noise (\ref{eq:linearVanderModel}). Hence, the RMSEs achieving the correponding RCRBs  validates our noise approximation discussed in Appendix \ref{ap:approxNoiseModel} for the given experimental setup. For the linear model (\ref{eq:linearVanderModel}), the proposed solution is the minimum variance unbiased estimator under Gaussian noise assumption. In this simulation, without loss of generality, we assume that the order of approximation $L$ is known. Alternatively, iterative solutions such as iMGLS \cite{rajanJ1} can be employed to estimate $L$. For a detailed discussion on the effect of $L$ on the distance estimation, particularly for an asynchronous network, see \cite{rajanJ1}. 
\color{black}


\begin{figure*} \centering 
\centering
{   
    \subfloat{
        \psfrag{xT}[cb]	{\small (a) Relative position: $\bXu$ [m]}
		\psfrag{xX}[cc]	{\small No. of Two-way comms. ($K$)}
		\psfrag{xY}[cb]	{\small RMSE}
		\psfrag{xLa}[lb]{\small MDS}
		\psfrag{xLb}[lb]{\small RCRB-Unconstrained}	
    \includegraphics[width=0.28\textwidth]{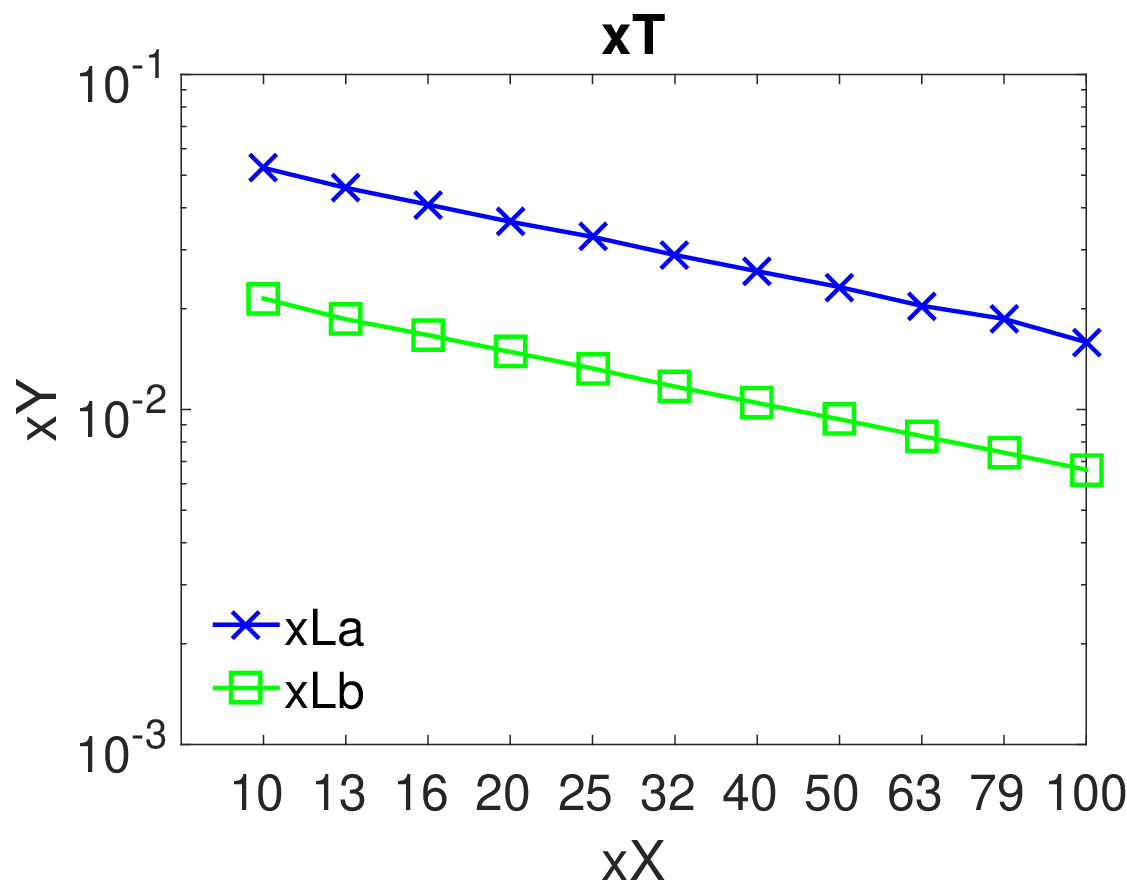}}
}
\;
{
    \subfloat{
        \psfrag{yT}[cb]	{\small (b) Relative velocity: $\bYu_1$ [ms$^{-1}$]}
		\psfrag{yX}[cc]	{\small No. of Two-way comms. ($K$)}
		\psfrag{yY}[cb]	{\small  }
		\psfrag{yLa}[lb]{\small LLS}
		\psfrag{yLb}[lb]{\small WLLS}
		\psfrag{yLc}[lb]{\small RCRB-Constrained}		
		\psfrag{yLd}[lb]{\small RCRB-Unconstrained}	
    \includegraphics[width=0.28\textwidth]{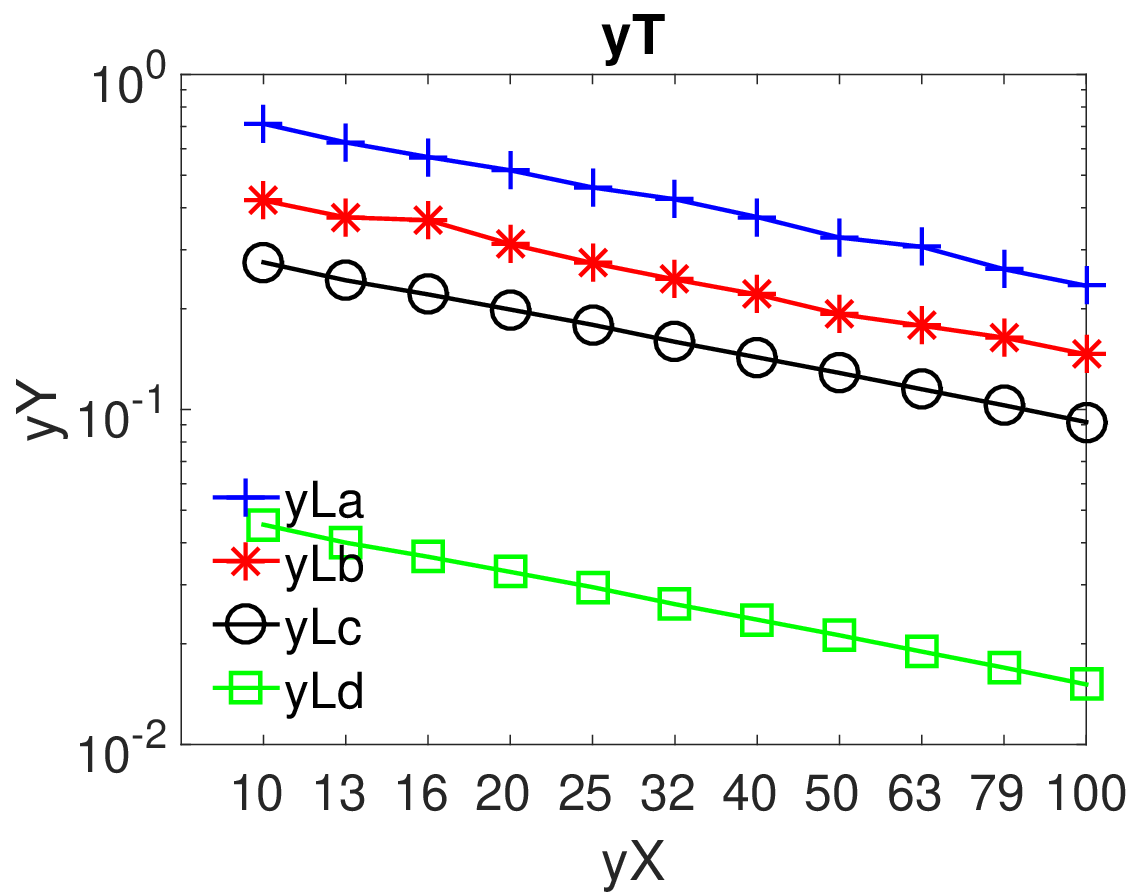}}
}
\;
{
    \subfloat{
        \psfrag{zT}[cb]	{\small (c) Relative acceleration $\bYu_2$ [ms$^{-2}$]}
		\psfrag{zX}[cc]	{\small No. of Two-way comms. ($K$)}
		\psfrag{zY}[cc]	{\small  }
		\psfrag{zLa}[lb]{\small LLS}
		\psfrag{zLb}[lb]{\small WLLS}
		\psfrag{zLc}[lb]{\small RCRB-Constrained}		
		\psfrag{zLd}[lb]{\small RCRB-Unconstrained}	
    \includegraphics[width=0.28\textwidth]{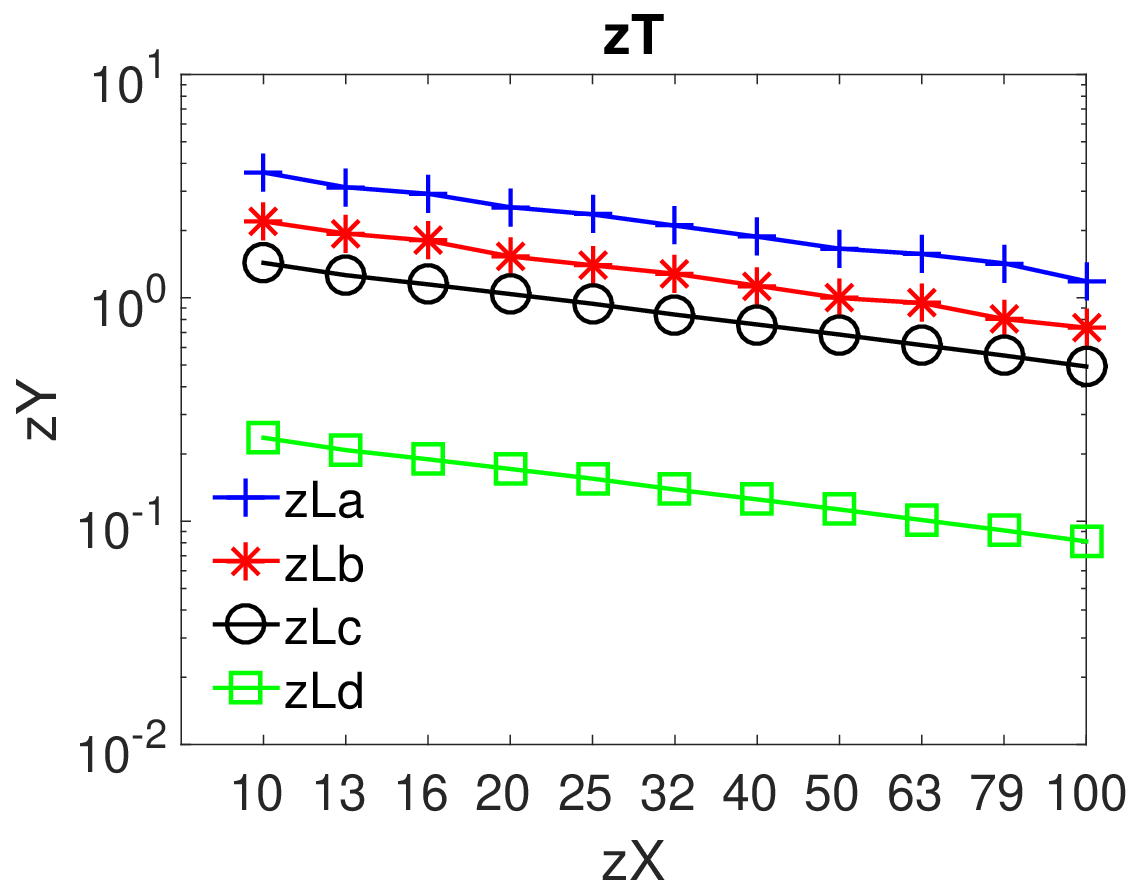}}
}
\\ \vspace{5mm}
{
    \subfloat{		
        \psfrag{xST}[cb] {\small (d) Relative position: $\bXu$ [m]}
		\psfrag{xSX}[cc] {\small $-10\log_{10}(\sigma)$ [dB meters]}
		\psfrag{xSY}[cb] {\small RMSE }
		\psfrag{xSLa}[lb]{\small MDS}
		\psfrag{xSLb}[lb]{\small RCRB-Unconstrained}		
    \includegraphics[scale=0.28]{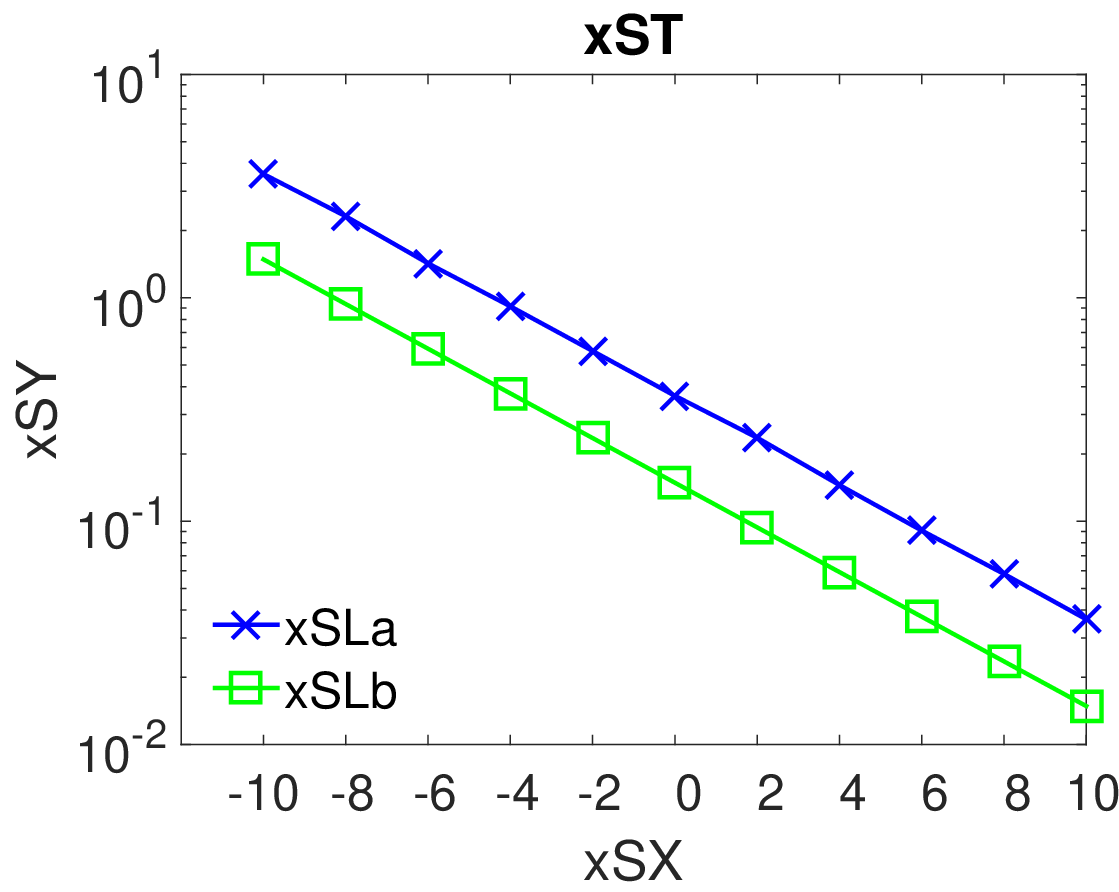}}
}
\;
{
    \subfloat{		
        \psfrag{yST}[cb]	{\small (e) Relative velocity: $\bYu_1$ [ms$^{-1}$]}
		\psfrag{ySX}[cc]	{\small $-10\log_{10}(\sigma)$ [dB meters]}
		\psfrag{ySY}[cb]	{\small }
		\psfrag{ySLa}[lb]{\small LLS}
		\psfrag{ySLb}[lb]{\small WLLS}
		\psfrag{ySLc}[lb]{\small RCRB-Constrained}
		\psfrag{ySLd}[lb]{\small RCRB-Unconstrained}
    \includegraphics[scale=0.28]{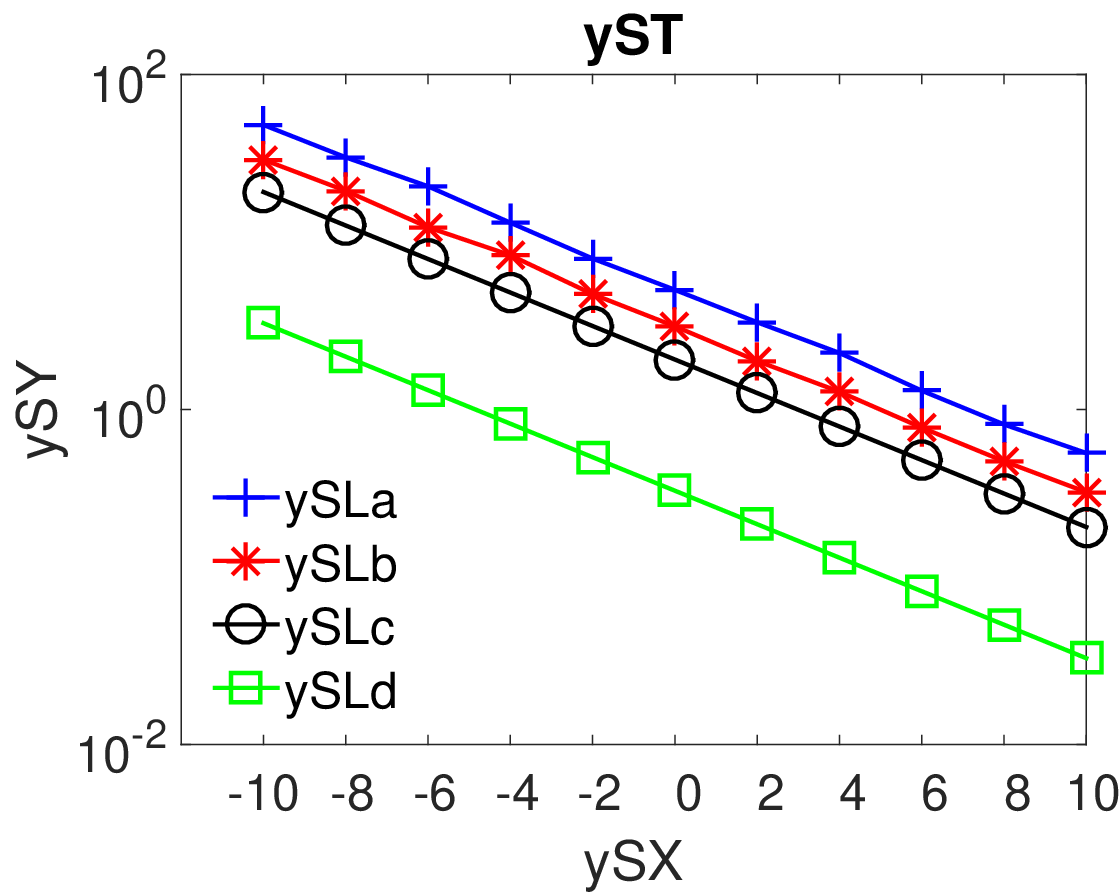}}
}
\;
{
    \subfloat{
        \psfrag{zST}[cb]	{\small (f) Relative acceleration $\bYu_2$ [ms$^{-2}$]}
		\psfrag{zSX}[cc]	{\small $-10\log_{10}(\sigma)$ [dB meters]}
		\psfrag{zSY}[ct]	{\small }
		\psfrag{zSLa}[lb]{\small LLS}
		\psfrag{zSLb}[lb]{\small WLLS}
		\psfrag{zSLc}[lb]{\small RCRB-Constrained}		
		\psfrag{zSLd}[lb]{\small RCRB-Unconstrained}		
    \includegraphics[scale=0.28]{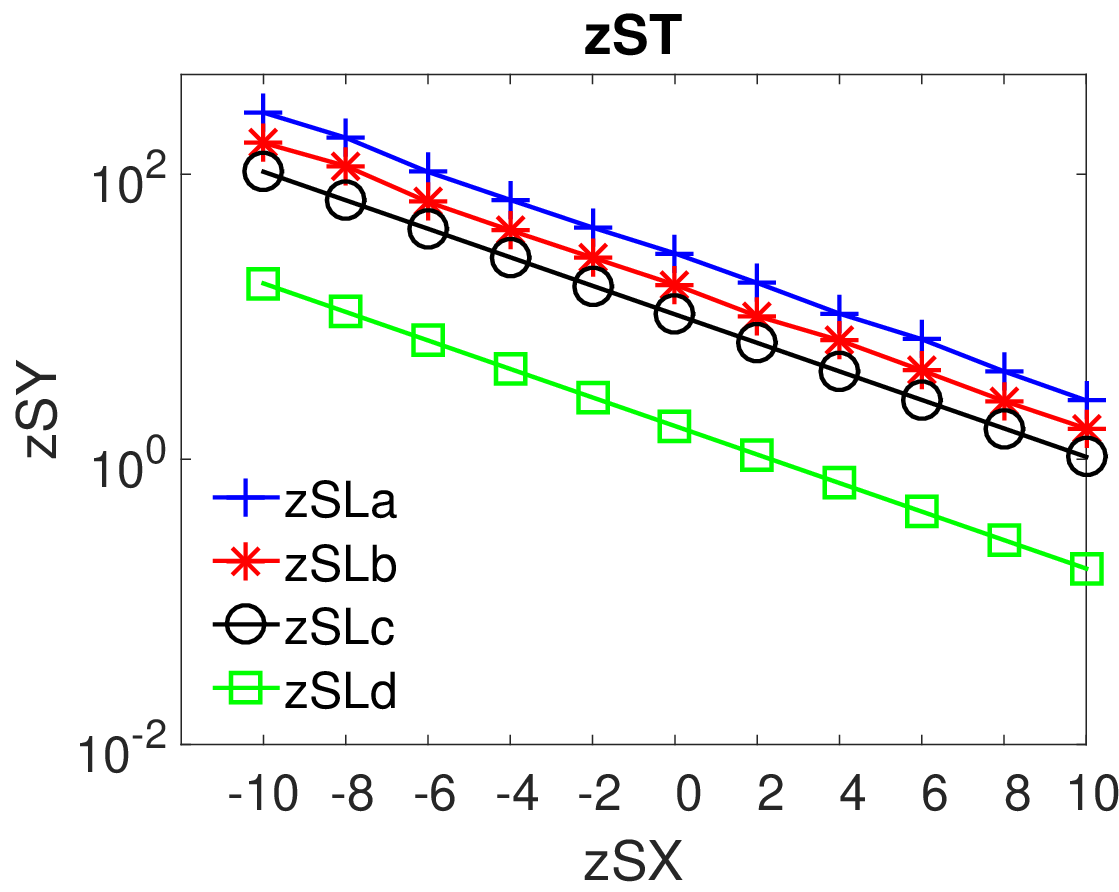}}
} \vspace{0.5\baselineskip}	
  \caption{\small \emph{\textbf{Relative Kinematics:}} \textsl{\textbf{Varying $K$:}} RMSEs (and RCRBs) of (a) Relative position ($\bXu$), (b) Relative velocity ($\bYu_1$) and (c) Relative acceleration ($\bYu_2$) for varying number of communications ($K$) between the $N=10$ \emph{mobile} nodes for $\sigma=0.1$ meters. \textsl{\textbf{Varying $\sigma$:}} RMSEs (and RCRBs) of (d) Relative position ($\bXu$), (e) Relative velocity ($\bYu_1$) and (f) Relative acceleration ($\bY_2u$), for a network of $N=10$ exchanging $K=10$ timestamps, where the Noise on the time markers ($\sigma$) is varied.}
  \label{fig:relativeKinematicsSim}
\end{figure*}

\subsection{Relative kinematics} The estimated relative range parameters yield the desired relative kinematics matrices. \figurename~\ref{fig:relativeKinematicsSim} shows the RMSEs (and RCRBs) of all the relative kinematic estimates. The MDS-based relative position estimates presented in \figurename~\ref{fig:relativeKinematicsSim}(a) and \figurename~\ref{fig:relativeKinematicsSim}(d), perform well against the derived oracle-bound, which was also observed in \cite{rajanJ1}. In case of the relative velocity and acceleration, we assume the minimal constraint $\bar{\bC}_1$ for analysis. Note that the unconstrained oracle-bounds are lower as compared to the CCRB, for a fixed SNR and increasing K. The WLLS solution outperforms the LLS solutions for both velocity and acceleration estimation, and asymptotically achieve the derived respective CCRBs. 

\begin{figure*} \centering 
\centering
{   
    \subfloat{
		\psfrag{yT}[cb]	{\small }
		\psfrag{yX}[cc]	{\small No. of Two-way comms. ($K$)}
		\psfrag{yY}[cb]	{\small RMSE }
		\psfrag{yLa}[lb]{\small LMDS}
		\psfrag{yLb}[lb]{\small WLLS}
		\psfrag{yLc}[lb]{\small RCRB-Constrained}		
		\psfrag{yLd}[lb]{\small RCRB-Unconstrained}		
        \includegraphics[scale=0.28]{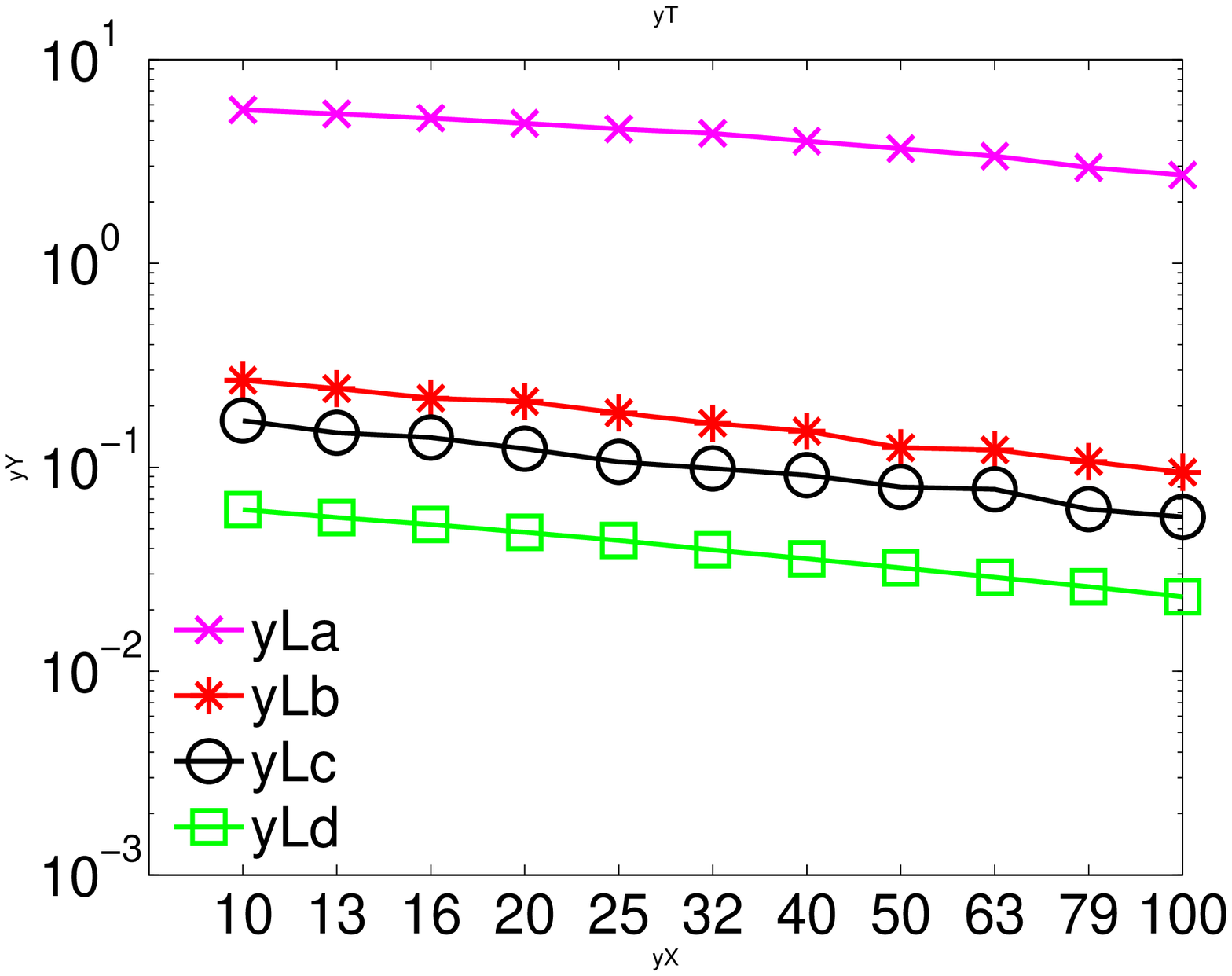}}
}
\;
{
    \subfloat{
        \psfrag{yST}[cb]	{\small }
		\psfrag{ySX}[cc]	{\small $-10\log_{10}(\sigma)$ [dB meters]}
		\psfrag{ySY}[cb]	{\small RMSE}
		\psfrag{ySLa}[lb]{\small LMDS}
		\psfrag{ySLb}[lb]{\small WLLS}
		\psfrag{ySLc}[lb]{\small RCRB-Constrained}
		\psfrag{ySLd}[lb]{\small RCRB-Unconstrained}
        \includegraphics[scale=0.28]{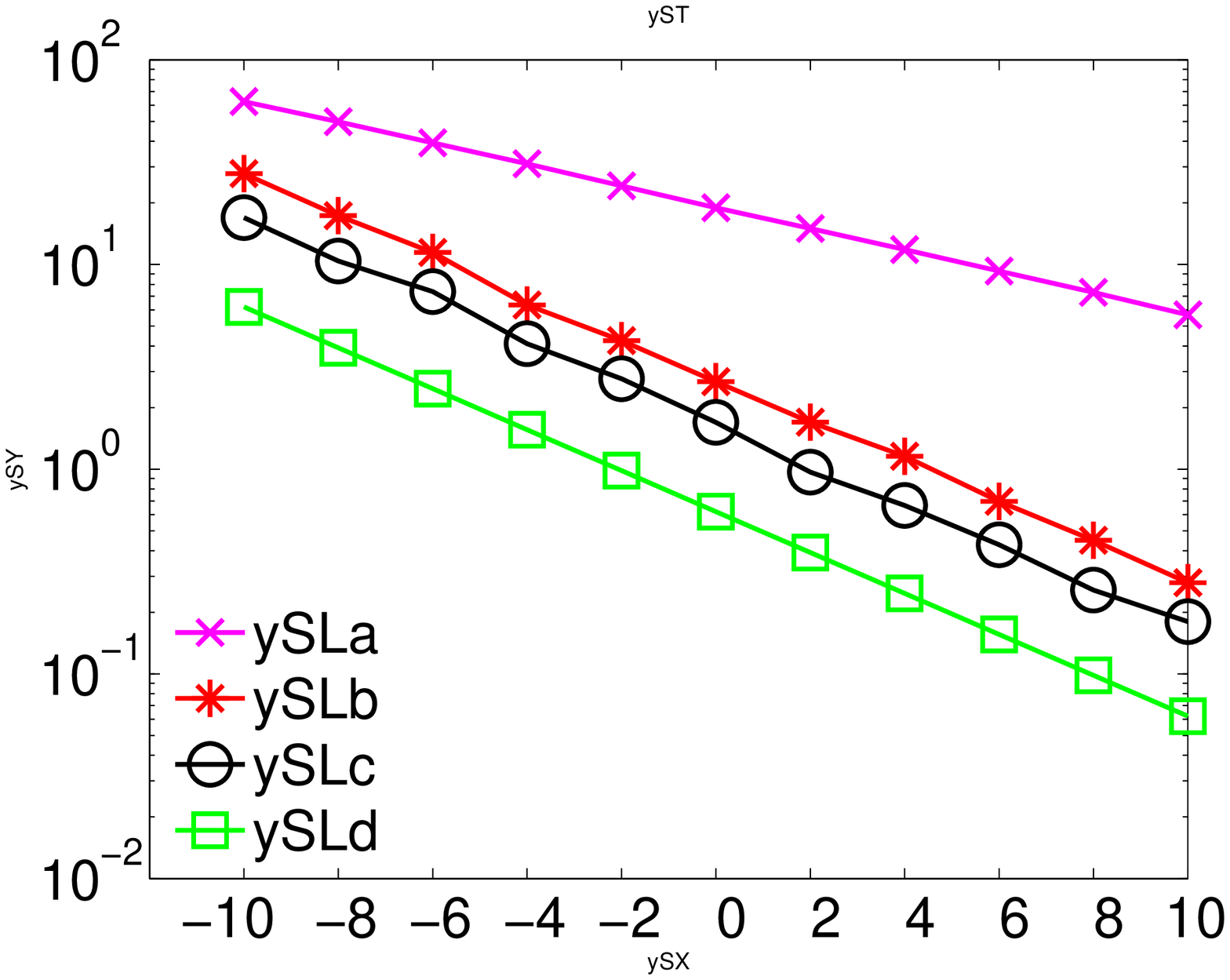}}
}
	\caption{\small \emph{\textbf{Comparison of relative velocity estimators:}} RMSEs (and RCRBs) of relative range parameters $\bY_1$ for varying number of communications ($K$) for $\sigma=0.1$ meters (top) and varying $\sigma$ (bottom) between the $N=10$ \emph{mobile} nodes.}
\label{fig:constantVelSim}
\end{figure*}

To compare the performance of the proposed relative velocity estimator against the MDS-based relative velocity estimation (\ref{eq:velMDS}), we perform another experiment. The MDS-based algorithm for relative velocity estimation assumes the nodes are in linear motion. Hence, we set $\bY_2= \bzero_{P,N}$ in (\ref{eq:posVelAccValues}) and re-implement the dynamic ranging algorithm for $L=2$ and plot the standard deviation of the estimates in \figurename~\ref{fig:constantVelSim}. Under the constant velocity assumption, the CCRB is comparable to the oracle-bound. The proposed WLLS solution outperforms the MDS-based estimator, especially for higher SNR and lower number of pair-wise communications. This is perhaps not surprising, since the MDS-based estimator relies on all the $\bR, \bdR, \bddR$ where the noise variance on these regression coefficients typically increase with the range-order for a Taylor basis (see \figurename~\ref{fig:rangeSim}). In comparison, the WLLS solution is dependent only on range $\bR$ and range rates $\bdR$. 

\subsection{Absolute kinematics} 
\begin{figure*}
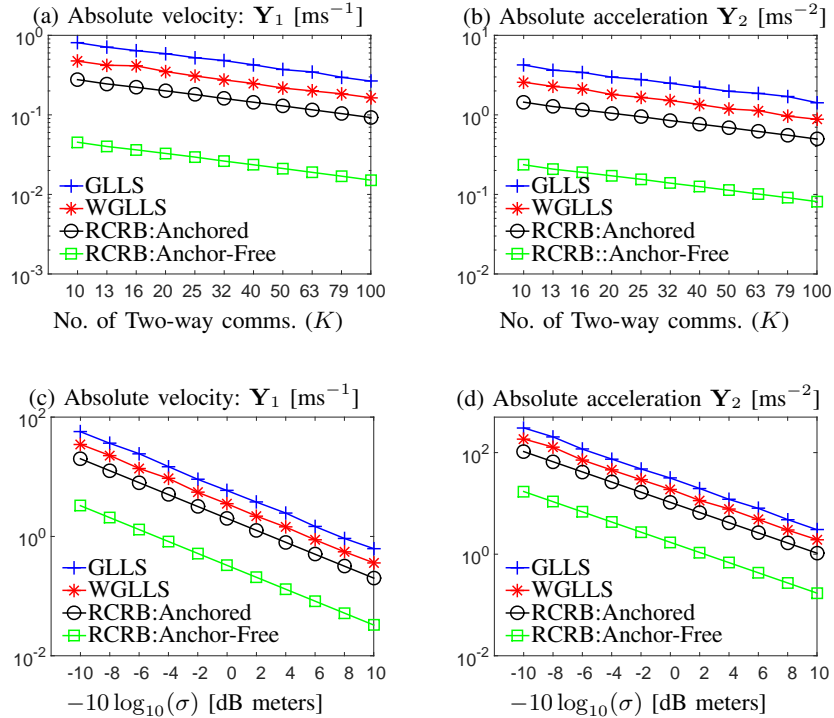
 \centering 
\centering
{   
    \subfloat{
		\psfrag{yT}[cb]	{\small (a) Absolute velocity: $\bY_1$ [ms$^{-1}$]}
		\psfrag{yX}[cc]	{\small No. of Two-way comms. ($K$)}
		\psfrag{yY}[cb]	{\small  }
		\psfrag{yLa}[lb]{\small GLLS}
		\psfrag{yLb}[lb]{\small WGLLS}
		\psfrag{yLc}[lb]{\small RCRB:Anchored}
		\psfrag{yLd}[lb]{\small RCRB:Anchor-Free}	
        \includegraphics[scale=0.28]{abs_vel.eps}}
}
\;
{
    \subfloat{
		\psfrag{zT}[cb]	{\small (b) Absolute acceleration $\bY_2$ [ms$^{-2}$]}
		\psfrag{zX}[cc]	{\small No. of Two-way comms. ($K$)}
		\psfrag{zY}[ct]	{\small  }
		\psfrag{zLa}[lb]{\small GLLS}
		\psfrag{zLb}[lb]{\small WGLLS}
		\psfrag{zLc}[lb]{\small RCRB:Anchored}
		\psfrag{zLd}[lb]{\small RCRB::Anchor-Free}
        \includegraphics[scale=0.28]{abs_acc.eps}}
}
\\ \vspace{5mm}
{
    \subfloat{
  		\psfrag{yST}[cb]	{\small (c) Absolute velocity: $\bY_1$ [ms$^{-1}$]}
		\psfrag{ySX}[cc]	{\small $-10\log_{10}(\sigma)$ [dB meters]}
		\psfrag{ySY}[cb]	{\small }
		\psfrag{ySLa}[lb]{\small GLLS}
		\psfrag{ySLb}[lb]{\small WGLLS}
		\psfrag{ySLc}[lb]{\small RCRB:Anchored}
		\psfrag{ySLd}[lb]{\small RCRB:Anchor-Free}
        \includegraphics[scale=0.28]{abs_velS.eps}}
}
\;
{
    \subfloat{
		\psfrag{zST}[cb]	{\small (d) Absolute acceleration $\bY_2$ [ms$^{-2}$]}
		\psfrag{zSX}[cc]	{\small $-10\log_{10}(\sigma)$ [dB meters]}
		\psfrag{zSY}[ct]	{\small }
		\psfrag{zSLa}[lb]{\small GLLS}
		\psfrag{zSLb}[lb]{\small WGLLS}
		\psfrag{zSLc}[lb]{\small RCRB:Anchored}
		\psfrag{zSLd}[lb]{\small RCRB:Anchor-Free}
        \includegraphics[scale=0.28]{abs_accS.eps}}
}
  \caption{\small \emph{\textbf{Absolute Kinematics:}} \textsl{\textbf{Varying $K$:}} RMSEs (and RCRBs) of (a) Absolute velocity ($\bY_1$) and (b) Absolute acceleration ($\bY_2$) for varying number of communications ($K$) between the $N=10$ \emph{mobile} nodes for $\sigma=0.1$ meters. \textsl{\textbf{Varying $\sigma$:}} RMSEs (and RCRBs) of (c) Absolute velocity ($\bY_1$) and (d) Absolute acceleration ($\bY_2$), for a network of $N=10$ nodes exchanging $K=10$ timestamps, where the noise on the time markers ($\sigma$) is varied.}
\label{fig:absoluteKinematicsSim}
\end{figure*}

\figurename~\ref{fig:absoluteKinematicsSim} shows the RMSEs and the corresponding RCRBs of the absolute velocity $\bY_1$ and acceleration $\bY_2$. We assume constraint (\ref{eq:constraintKnown}) to solve the proposed GLLS (\ref{eq:cclsAbs}) and WGLLS (\ref{eq:wclsAbs}) algorithms. The proposed estimators are seen to converge asymptotically to the derived CCRBs, while the CCRB itself is an order higher than the theoretical oracle-bound. The performance of the absolute kinematics is very similar to that of the relative kinematics (see \figurename~\ref{fig:relativeKinematicsSim}), which is due to the fact that the FIMs in both scenarios are dominated by the singular values of the relative position matrix.

\subsection{Relative and absolute time-varying positions} 
\begin{figure*}
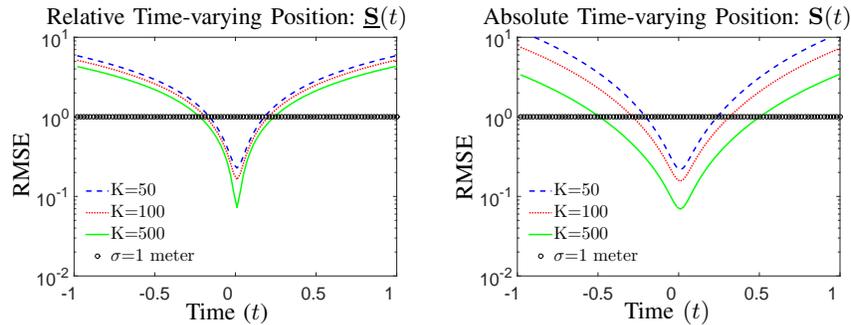
 
\centering
{   
    \subfloat{
		\psfrag{rT}[cb]	{\small Relative Time-varying Position: $\bSu(t)$}
		\psfrag{rX}[cb]	{\small Time ($t$)}
		\psfrag{rY}[cb]	{\small RMSE}
		\psfrag{rLa}[lb]{\small $K=50$}
		\psfrag{rLb}[lb]{\small $K=100$}
		\psfrag{rLc}[lb]{\small $K=500$}
		\psfrag{rLd}[lb]{\small $\sigma = 1 meter$}	
        \includegraphics[scale=0.28]{Xk_rel.eps}}
}
\;
{
    \subfloat{
		\psfrag{aT}[cb]{\small  Absolute Time-varying Position: $\bS(t)$}
		\psfrag{aX}[cb]{\small  Time $(t)$}
		\psfrag{aY}[cb]{\small  RMSE}
		\psfrag{aLa}[lb]{\small $K=50$}
		\psfrag{aLb}[lb]{\small $K=100$}
		\psfrag{aLc}[lb]{\small $K=500$}
		\psfrag{aLd}[lb]{\small $\sigma = 1 meter$}	
    \includegraphics[scale=0.28]{Xk_abs.eps}}
}
	\caption{\small \emph{\textbf{Position over time:}} RMSEs of relative position $\underline{\bS}(t)$ and absolute position $\bS(t)$ over time for $K=[50, 100, 500]$ communications between a cluster of $N=10$ \emph{mobile} nodes, with $\sigma=1$ meter.}
\label{fig:posOverTime}
\end{figure*}

The estimation of the node kinematics enable us to reconstruct the time-varying relative positions $\underline{\bS}(t)$ and time-varying absolute positions $\bS(t)$, from (\ref{eq:relPos_estimate}) and (\ref{eq:absPos_estimate}) respectively. We conduct experiments to study the effect of the proposed estimators on the time-varying positions. The RMSE plot for the absolute and relative time-varying positions around the region of interest at $t_0=0$ are shown in \figurename~\ref{fig:posOverTime}, where the number of communications are varied as $K=[50,100,500]$ with a Gaussian noise on the distance of $\sigma=1$ meter. For $K=500$, the RMSE estimate of both the relative and absolute position around $t_0$ shows an improvement by an order magnitude in comparison to the noise on the distance measurement, for the given experimental setup. This gain is primarily contributed during dynamic ranging, where $K$ data points are averaged using the Taylor basis which yields a factor $\sqrt{K}$ improvement on the estimate of the range parameters. Secondly, the performance deteriorates as we move away from $t_0$, which is a typical characteristic of the Taylor approximation. However, if Doppler measurements are available for radial velocities and other higher order derivatives, then the standard deviation of the estimators can be further reduced.
 
\begin{figure*}
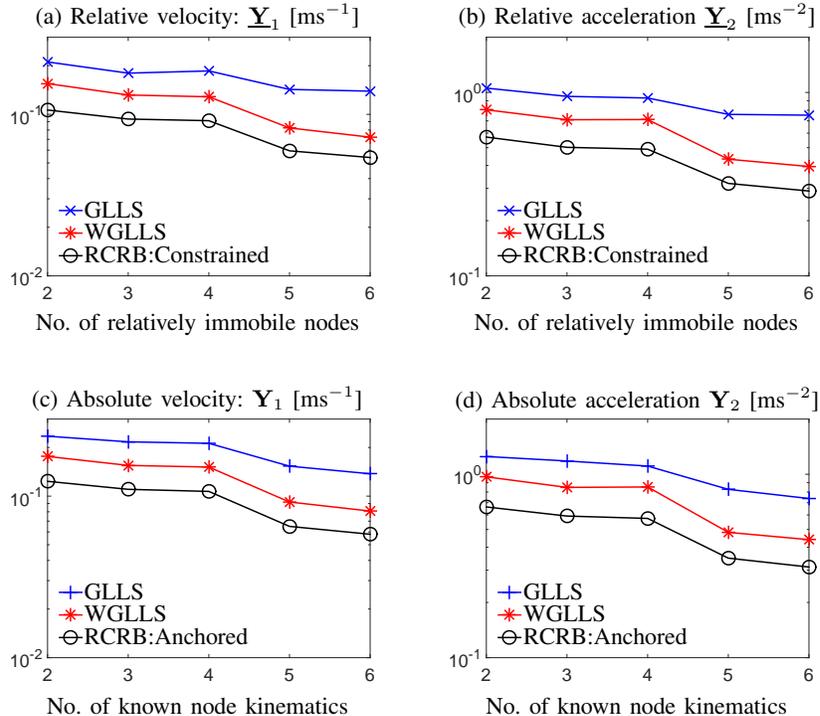
 \centering 
\centering
{   
    \subfloat{
		\psfrag{yT}[cb]	{\small (a) Relative velocity: $\bYu_1$ [ms$^{-1}$]}
		\psfrag{yX}[cc]	{\small No. of relatively immobile nodes}
		\psfrag{yY}[cb]	{\small  }
		\psfrag{yLa}[lb]{\small GLLS}
		\psfrag{yLb}[lb]{\small WGLLS}
		\psfrag{yLc}[lb]{\small RCRB:Constrained}
    \includegraphics[scale=0.28]{rel_vel_constraint.eps}}
}
\;
{
    \subfloat{
		\psfrag{zT}[cb]	{\small (b) Relative acceleration $\bYu_2$ [ms$^{-2}$]}
		\psfrag{zX}[cc]	{\small No. of relatively immobile nodes}
		\psfrag{zY}[ct]	{\small  }
		\psfrag{zLa}[lb]{\small GLLS}
		\psfrag{zLb}[lb]{\small WGLLS}
		\psfrag{zLc}[lb]{\small RCRB:Constrained}
    \includegraphics[scale=0.28]{rel_acc_constraint.eps}}
}
\\ \vspace{5mm}
{
    \subfloat{
		\psfrag{yT}[cb]	{\small (c) Absolute velocity: $\bY_1$ [ms$^{-1}$]}
		\psfrag{yX}[cc]	{\small No. of known node kinematics}
		\psfrag{yY}[ct]	{\small  }
		\psfrag{yLa}[lb]{\small GLLS}
		\psfrag{yLb}[lb]{\small WGLLS}
		\psfrag{yLc}[lb]{\small RCRB:Anchored}
		\psfrag{yLd}[lb]{\small RCRB:Anchor-Free}
    \includegraphics[scale=0.28]{abs_vel_constraint.eps}}
}
\;
{
    \subfloat{
		\psfrag{zT}[cb]	{\small (d) Absolute acceleration $\bY_2$ [ms$^{-2}$]}
		\psfrag{zX}[cc]	{\small No. of known node kinematics}
		\psfrag{zY}[ct]	{\small  }
		\psfrag{zLa}[lb]{\small GLLS}
		\psfrag{zLb}[lb]{\small WGLLS}
		\psfrag{zLc}[lb]{\small RCRB:Anchored}
		\psfrag{zLd}[lb]{\small RCRB:Anchor-Free}
        \includegraphics[scale=0.28]{abs_acc_constraint.eps}}
}
  \caption{\small \emph{\textbf{Effect of increasing constraints:}} \textsl{\textbf{Relative kinematics:}} RMSEs (and RCRB) of (a) Relative velocity ($\bYu_1$) and (b) Relative acceleration ($\bYu_2$) for varying number of relatively immobile nodes. \textsl{\textbf{Absolute kinematics:}} RMSEs (and RCRBs) of (c) Absolute velocity ($\bY_1$) and (d) Absolute acceleration ($\bY_2$) for varying number of known node kinematics}
\label{fig:choiceOfConstraints}
\end{figure*}

\subsection{Choice of constraints} In the previous sections, we evaluated the proposed algorithms under minimal constraints. Now, we perform experiments to understand the effect of incorporating additional constraints (or references) on the performance of the proposed estimators. These additional constraints implicitly reduce the parameter subspace, and consequently affect the overall RMSE of the proposed estimators. In order to understand this variation, we set  $N_z=1$ in our performance metric (\ref{eq:rmseAbs}) for the following simulations. To estimate the relative kinematics in a $2$ dimensional scenario, a unique solution is feasible if at least $2$ nodes are relatively immobile (see \appendixname\ \ref{ap:underdeterminedLyapunov}). If more nodes are immobile, then the constraints in (\ref{eq:constraintImmobile}) can be readily extended to incorporate this supplementary information. Similarly, in case of absolute velocity and acceleration estimation,  a minimum of at least $2$ node kinematics must be known. Therefore, in the following experiments we vary the number of known kinematics (or immobile nodes) from $2$ to $6$, for a fixed number of two way communications $K=100$ with $\sigma=0.1$ meters. \figurename\ \ref{fig:choiceOfConstraints} shows the results of the GLLS and WGLLS algorithms for estimating the absolute and relative kinematics, along with the respective CCRBs. Not surprisingly, we observe an improvement in the performance of the algorithms with the additional constraints. In addition, unlike the GLLS estimator, the WGLLS estimator asymptotically achieves the respective CCRBs.

%
%
%
%
%
%
%
%
%
%

\section{Conclusions} Understanding the relative kinematics of an anchorless network of mobile nodes is paramount for reference-free localization technologies of the future. We presented a novel data model which relates the time-varying distance measurements to the $M$th order relative kinematics for an anchorless network of mobile nodes. The derived data model takes the form of a Lyapunov-like equation, which under certain constraints, can be recursively solved for estimating the relative velocity, acceleration and higher order derivatives. Closed form constrained estimators, such as the LS and WLS are proposed, which are also the BLUE for the given data model. \Cramer-Rao lower bounds are derived for the new data model and the performance of the proposed algorithms is validated using simulations. Although our focus is on relative localization, the proposed model and solutions can be broadly applied to understand feature variations in Euclidean space, with applications in general exploratory data analysis.

In our future work, we are keen in addressing two research challenges. Firstly, our focus in this article has been on finding unique solutions to time-derivatives of the relative position matrix. To this end, unbiased constrained estimators are proposed to solve the under-determined Lyapunov-like equation. However, more generally, regularized algorithms can be employed, such as Ridge regression \cite{golub1999}, subset selection \cite{lawson74}  or Lasso \cite{tibshirani1996},  without the need for equality constraints on the cost function. The estimates of such unconstrained algorithms can be corroborated against the unconstrained \Cramer-Rao bound derived in this article. Furthermore, the algorithms are inherently centralized in nature, which could be distributed for resource constrained implementation. Finally, the proposed framework is particularly helpful for cold-start scenarios when there is no apriori information on the position or higher order kinematics. In practice, given the cold-start solution on relative velocity and higher order kinematics, a state-space model readily emerges for dynamic tracking of the relative positions over time, which can be elegantly solved using adaptive filters.

\appendices

\section{Approximate noise model} \label{ap:approxNoiseModel} To estimate the range parameters from time-varying propagation delays, we presented the dynamic ranging model in (\ref{eq:linearVanderModel}), with additive Gaussian noise \ie \begin{equation} \bV\btheta = \btau + \bEta, 
\label{eq:globalNormal_crb}
\end{equation} where $\bV$ is the Vandermonde-like matrix, $\btheta$ contains the unknown range coefficients, $\btau$ contains all the propagation delays and  $\bEta$ is the noise vector plauging the propagation delays. In practise, the noise is on the time markers and subsequently on the Vandermonde matrix. However, under certain nominal assumptions, the above model is valid, which we discuss in this section.

We begin with the noiseless pairwise time-varying dynamic ranging model, which we recollect from (\ref{eq:tauij_dij_t_model_k}) as below \begin{equation}
\underline{r}_{ij} + 
\underline{\dot{r}}_{ij} \Delta T_k +
\underline{\ddot{r}}_{ij}\Delta T_k^2+ \hdots
= |T_{ij,k} - T_{ji,k}|
= \tau_{ij,k} , 
\end{equation} where we introduce $\Delta T_k=(T_{ij,k}-T_0)$ for notational simplicity. In reality, there is noise plaguing the time markers and hence we have, \begin{equation}
\underline{r}_{ij} + 
\underline{\dot{r}}_{ij}(\Delta T_k + \eta_{i,k}) + 
\underline{\ddot{r}}_{ij}(\Delta T_k  + \eta_{i,k}) ^2+ 
\hdots = \tau_{ij,k} + \eta_{ij,k}, 
\end{equation} where $\{\eta_{i,k}, \eta_{j,k}\}$ are the noise terms on the time markers at node $i$ and node $j$ respectively, and $\eta_{ij,k}= \eta_{i,k}-\eta_{j,k}$ is the pairwise noise error of the node pair $(i,j)$. Expanding the polynomial and rearranging the terms, we have \begin{equation}
\underline{r}_{ij} + 
\underline{\dot{r}}_{ij}\Delta T_k + 
\underline{\ddot{r}}_{ij}\Delta T_k^2+ 
\hdots + \bar{\eta}_{i,k}
= \tau_{ij,k} + \eta_{ij,k}.
\end{equation} Here $\bar{\eta}_{i,k}$ is the cumulative noise error from the Taylor approximation, which is expressed as \begin{equation}
\bar{\eta}_{i,k} =  
\eta_{i,k}\Big( 
\dot{\ur}_{ij,k} +
2\ddot{\ur}_{ij,k}\Delta T_k + 
\hdots \Big) + 
\eta^2_{i,k}\Big(\ddot{\ur}_{ij,k} + \hdots \Big) + \hdots \approx\ 0,
\label{eq:noise_approximation}
\end{equation} and approximated to $0$. This approximation is valid under two assumptions. Firstly, we assume that the time stamps are measured with high SNR, \ie we consider standard deviations of $\le 10^{-7}$ seconds on the time stamps, which is necessary to achieve meter level accuracies is conventional two-way ranging based localization solutions \cite{patwari05,simonetto2014distributed}. As a consequence, we ignore the second order noise term $\eta^2_{i,k}$, and other higher order noise terms in (\ref{eq:noise_approximation}). Secondly, observe from definition (\ref{eq:rangeTranslation}) that the coefficients $\{\dot{\ur}, \ddot{\ur}, \hdots \}$ are scaled by $c^{-1}$, where $c=3 \times 10^8$ m/s for free space. Therefore, the Taylor coefficients are significantly small and subsequently, the term  $(\dot{\ur}_{ij,k} + 2\ddot{\ur}_{ij,k}\Delta T_{k} + \hdots)$ is negligible for a measurement period of upto a few seconds. This is a pragmatic assumption, since we are only interested in the instantaneous relative kinematics of the nodes around a small time interval. In summary, for small measurement periods in high SNR scenarios, the noise parameter $\bar{\eta}_{i,k} \approx 0$, and under these assumptions (\ref{eq:linearVanderModel}) holds.
\color{black}

\section{Underdetermined Lyapunov-like equation} \label{ap:underdeterminedLyapunov}

\newtheorem{theorem}{Theorem} \begin{theorem}[Underdetermined Lyapunov-like equation] Given $\bX \in \mathbb{R}^{P \times N}$ and $\bB \in \mathbb{R}^{N \times N}$ for $N > P$, the Lyapunov-like equation \begin{equation} \label{eq:apLyapunov}
\bX^T\bY + \bY^T\bX= \bB,
\end{equation} is rank-deficient by at least $\bar{P}= {P \choose 2}$.
\end{theorem} 

\begin{IEEEproof} Let the singular value decomposition of $\bX$ be \begin{equation} \bX= \bU_x \begin{bmatrix} \bLambda_x &  \bzero \end{bmatrix} \bV^T_x,
\end{equation} where $\bLambda_x \in \mathbb{R}^{P \times P}$ is a diagonal matrix containing the singular values and $\bU_x \in \mathbb{R}^{P \times P}, \text{and}\ \bV_x \in \mathbb{R}^{N \times N}$ are the corresponding singular vectors. Then, (\ref{eq:apLyapunov}) is \begin{equation}
\begin{bmatrix} \bLambda_x &  \bzero \end{bmatrix}^T\tilde{\bY} + 
\tilde{\bY}^T\begin{bmatrix} \bLambda_x &  \bzero \end{bmatrix} =
\tilde{\bB},
\end{equation} where 
\begin{eqnarray}
\tilde{\bB}&=&	
\begin{bmatrix} \tilde{\bB}_{11} & \tilde{\bB}_{12} \\  \tilde{\bB}^T_{12} & \tilde{\bB}_{22} \end{bmatrix} =\ \bV^T_x\bB_x\bV_x, \\
\tilde{\bY} &=& 
\begin{bmatrix} \tilde{\bY}_{1} & &\tilde{\bY}_2 \end{bmatrix}=\ \bU^T_x\bY\bV_x,
\end{eqnarray} where $\tilde{\bY}_{1} \in \mathbb{R}^{P \times P}$, $\tilde{\bY}_{2} \in \mathbb{R}^{P \times N-P}$ and $\tilde{\bB}_{22}= \bzero$ for the equation to be consistent. A solution to the system (\ref{eq:apLyapunov}) is obtained by solving for $\tilde{\bY}$ the set of equations, \begin{eqnarray} 
	\label{eq:tY1}
	\bLambda_x\tilde{\bY}_1 + \tilde{\bY}^T_1\bLambda_x &=& \tilde{\bB}_{11},\\
	\label{eq:tY2}
	\bLambda_x \tilde{\bY}_2  &=&\tilde{\bB}_{12}.
\end{eqnarray} 

An estimate for  $\tilde{\bY}_2$ is straightforward and is given by $\hat{\tilde{\bY}}_2 = \bLambda^{-1}_x \tilde{\bB}_{12}$. Let $\tilde{\bLambda}_x, \tilde{\bY}_1$ and $\tilde{\bB}_{11}$ be partitioned into \begin{equation}
\begin{bmatrix}
\sigma_{1}&		0 \\
0&						\bLambda_{x,1}
\end{bmatrix}, \quad
\begin{bmatrix}
y_{11}&							\tilde{\by}_{12} \\
\tilde{\by}_{21}&		\tilde{\bY}_{1,1}
\end{bmatrix}, \quad
\begin{bmatrix}
\tilde{b}_{11}&				\tilde{\bb}_{12} \\
\tilde{\bb}_{12}^T&		\tilde{\bB}_{11,1}
\end{bmatrix}, 
\end{equation} then (\ref{eq:tY1}) is equivalent to solving \begin{eqnarray}
y_{11}&=&	\tilde{b}_{11}/2\sigma_1, \label{eq:partition1}\\
\sigma_1\tilde{\by}_{12} + \tilde{\by}^T_{21}\bLambda_{x,1} &=&	\tilde{\bb}_{12}, \label{eq:partition2} \\
\bLambda_{x,1}\tilde{\bY}_{1,1} + \tilde{\bY}^T_{1,1}\bLambda_{x,1} &=& \tilde{\bB}_{11,1} \label{eq:partition3}.
\end{eqnarray} Note that the solution to $y_{11}$ in (\ref{eq:partition1}) is straightforward, however the solution to off-diagonal terms $\tilde{\by}_{12},\tilde{\by}_{21}$ is underdetermined. Furthermore, since (\ref{eq:partition3}) is in form similar to the (\ref{eq:tY1}), $\tilde{\bY}_{1,1}$ can be estimated recursively \cite{chu1989}. Thus, the diagonal terms of the $P$ dimensional matrix $\tilde{\bY}_{1,1}$ can be estimated, however to resolve the ambiguity of the off-diagonal terms atleast $\bar{P}={P \choose 2}$ constraints are required.
\end{IEEEproof} 

\section{Karush-Kuhn-Tucker (KKT) equations} \label{ap:KKT} A solution to minimize the equality constrained function of the form \begin{equation}
\min_{\by} \norm{ \bA\by- \bb \big}^2 \quad \text{s.t.}\quad \bC\by=\bd,
\end{equation} is obtained by solving the Karush-Kuhn-Tucker (KKT) system of equations, \begin{equation}
\label{eq:KKT_solution}
\begin{bmatrix}
\hat{\by} \\
\hat{\blambda}
\end{bmatrix} =
\begin{bmatrix}
2\bA^T\bA& \bC^T \\
\bC      							& \bzero_{N_2, N_2}\\
\end{bmatrix}^{-1}
\begin{bmatrix}
2\bA^T\bb \\
\bd
\end{bmatrix} ,
\end{equation} where $\hat{\by}$ is an estimate of the unknown parameter $\by$ and $\hat{\blambda}$ collects the corresponding Lagrange multipliers. The problem has a feasible solution provided $\begin{bmatrix} \bA \\ \bC \end{bmatrix}$ is full column rank \cite{boydConvexOptimization}. 

\section{Expression for $\bSigma_{b,M}$} \label{ap:covariance_B} 
We present an explicit expression for the covariance matrix $\bSigma_{b,M}$, which is obtained by ignoring higher order noise terms \ie for sufficiently large SNR. For $M=1$, \ie relative velocity, we have 
\begin{eqnarray}
\bSigma_{b,1}		&\approx& \tilde{\bP}\Big(\bPsi_r\bar{\bSigma}_{\dr}\bPsi_r + \bPsi_{\dr}\bar{\bSigma}_{r} \bPsi_{\dr}\Big)\tilde{\bP},
\end{eqnarray} and for $M=2$, \ie relative acceleration, we have \begin{eqnarray}
\bSigma_{b,2}		&\approx& \tilde{\bP}\Big(	\bPsi_r\bar{\bSigma}_{\ddr}\bPsi_r 
																+ \bPsi_{\ddr}\bar{\bSigma}_{r}\bPsi_{\ddr} 
																+ 4\bPsi_{\dr}\bar{\bSigma}_{\dr}\bPsi_{\dr}  \Big) \tilde{\bP} \nonumber \\
								&&\; +\ 4\bPsi_{y}\bar{\bSigma}_{\dot{x}}\bPsi_{y},
\end{eqnarray} where we  $\tilde{\bP}\triangleq\bP\otimes\bP$, $\bPsi_r\triangleq\ \text{diag}\big(\text{vec}\big(\bR)\big), \bPsi_{\dr}\triangleq\ \text{diag}\big(\text{vec}\big(\bdR)\big)$ and $\bPsi_{\ddr}\triangleq\ \text{diag}\big(\text{vec}\big(\bddR)\big)$. The matrix $\bPsi_{y} = \bA_{y,1}$ for absolute kinematics and $\bPsi_{y} = \bAu_{y,1}$ for relative kinematics. Observe that the diagonal elements of the range parameters $\bR, \dot{\bR}, \ddot{\bR}, \hdots $ contain zeros and consequentially the matrices $\bPsi_{r}, \bPsi_{\dr}, \bPsi_{\ddr}, \hdots$ are singular. Hence the covariance matrix $\bSigma_{b,M}$ is in general rank deficient. Furthermore, $\bA_y$ in (\ref{eq:Aydef}) is rank deficient by definition and subsequently $\bSigma_{\rho}$ (\ref{eq:rhoCov}) is ill-conditioned and therefore, we use the Moore-Penrose pseudo-inverse in (\ref{eq:unconstrainedFIM}) and (\ref{eq:rhoCovEstimate}). An expression for higher order $M > 2$ can be similarly derived.

\bibliographystyle{IEEEtran}
\bibliography{main}
\end{document}